\documentclass[preprint,aps,nofootinbib]{revtex4-1}
\pdfoutput=1

\usepackage{dcolumn}
\usepackage{bm}
\usepackage{epsfig}
\usepackage{graphicx}
\usepackage{float}

\usepackage{amsmath}
\usepackage{amsfonts}
\usepackage{amssymb}
\usepackage{multirow}
\usepackage{color}
\usepackage{xcolor}
\usepackage{adjustbox}
\usepackage{subcaption}
\usepackage[toc,page]{appendix}

\usepackage[utf8]{inputenc}
\usepackage{slashed}
\usepackage{indentfirst}
\usepackage{amsfonts}
\usepackage{mathtools}
\usepackage{hyperref} 
\usepackage{natbib} 
\usepackage{ulem}

\def\mtt{m_{\tilde{t}}}
\def\mttwo{m_{\tilde{t}_2}}
\def\mtone{m_{\tilde{t}_1}}
\def\mbone{m_{\tilde{b}_1}}
\def\tt{\tilde{t}}
\def\tone{\tilde{t}_1}
\def\ttwo{\tilde{t}_2}
\def\bt{\tilde{b}}
\def\bone{\tilde{b}_1}

\def\mcone{m_{\tilde{\chi}_1}}

\def\cc{\tilde{\chi}^\pm}
\def\cn{\tilde{\chi}}
\def\cone{\tilde{\chi}_1}
\def\ctwo{\tilde{\chi}_2}
\def\mq{m_{Q_3}}
\def\muu{m_{u_3}}
\def\md{m_{d_3}}

\begin{document}

\title{Second Stop and Sbottom Searches with a Stealth Stop}
\author{Hsin-Chia Cheng, Lingfeng Li, Qin Qin}
\affiliation{Department of Physics, University of California, Davis, California 95616, USA}

\begin{abstract}
The top squarks (stops) may be the most wanted particles after the Higgs boson discovery. The searches for the lightest stop have put strong constraints on its mass. However, there is still a search gap in the low mass region if the spectrum of the stop and the lightest neutralino is compressed. In that case, it may be easier to look for the second stop since naturalness requires both stops to be close to the weak scale. The current experimental searches for the second stop are based on the simplified model approach with the decay modes
$\ttwo \to \tone Z$ and $\ttwo \to \tone h$. 
However, in a realistic supersymmetric spectrum there is always a sbottom lighter than the second stop, hence the decay patterns are usually more complicated than the simplified model assumptions. In particular, there are often large branching ratios of the decays $\ttwo \to \bone W$ and $\bone \to \tone W$ as long as they are open. The decay chains can be even more complex if there are intermediate states of additional charginos and neutralinos in the decays. By studying several MSSM benchmark models at the 14 TeV LHC, we point out the importance of the multi-$W$ final states in the second stop and the sbottom searches, such as the same-sign dilepton and multilepton signals, aside from the traditional search modes. The observed same-sign dilepton excesses at LHC Run 1 and Run 2 may be explained by some of our benchmark models. 
We also suggest that the vector boson tagging and a new kinematic variable may help to suppress the backgrounds and increase the signal significance for some search channels. Due to the complex decay patterns and lack of the dominant decay channels, the best reaches likely require a combination of various search channels at the LHC for the second stop and the lightest sbottom.
\end{abstract}
\maketitle

\section{Introduction}
\label{sec:introduction}

Supersymmetry (SUSY) provides a most promising solution to the hierarchy problem to the standard model (SM). The quadratically divergent contributions to the Higgs mass-squared parameter from couplings to the SM fields are canceled by those of their superpartners. To keep the scale of the electroweak symmetry breaking natural, the superpartners are expected to have masses around or beneath the TeV scale. In particular, since the largest coupling to the Higgs in the SM is from the top quark, the superpartners of the top quark (top squarks or stops) play the most important role in canceling the quadratic divergence and are expected to be close to the weak scale in a natural theory.

On the other hand, the Higgs boson mass of 125 GeV also has important implications for the stop masses. In the minimal supersymmetric standard model (MSSM), the tree-level Higgs boson mass has an upper bound of $M_Z$. To get to 125 GeV, it requires large radiative contributions from the stop loops~\cite{Haber:1990aw,Okada:1990vk,Ellis:1990nz}. This could happen if the stops are heavy and/or the trilinear $A_t$ term of the stop sector is large~\cite{Hall:2011aa,Arbey:2011ab,Draper:2011aa,Carena:2011aa,Akula:2011aa}. To keep the fine-tuning minimal, it is preferable to have a large $A_t$ term so that the stops masses can be lowered to $\sim 1$ TeV or below. A large $A_t$ term implies large off-diagonal masses of the stop mass matrix so there will be a substantial mixing between the left-handed and the right-handed stops in the mass eigenstates. As a consequence, there will also be a sizable mass difference between the two stop mass eigenstates. 

As a key to the naturalness problem, the stops have been extensively searched for at the LHC. Assuming that the lightest neutralino $\tilde{\chi}^0$ is the lightest supersymmetric particle (LSP) and is stable, the search limit for the $\tilde{t} \to t \tilde{\chi}^0$ decay (assuming 100\% branch fraction) has reached $\sim 860$ GeV for $m_{\tilde{\chi}^0} \lesssim 250$~GeV at the current Run 2 of LHC~\cite{ATLAS:2016jaa,ATLAS:2016ljb,CMS:2016xva,CMS:2016vew,CMS:2016hxa,CMS:2016inz}.
From the naturalness point view, some fine-tuning is already required if the lightest stop is heavier than 860 GeV. However, the search limits are significantly weakened in the compressed region, where $m_{\tilde{t}_1}- m_{\tilde{\chi}^0} \lesssim m_t$. In particular, there is a gap along $m_{\tilde{t}_1}- m_{\tilde{\chi}^0} \approx m_t$ in the $m_{\tilde{t}_1}- m_{\tilde{\chi}^0}$ plane. In this case the top quark and the neutralino from the stop decay are roughly static in the stop rest frame. For the stop pair production, the neutralinos travel along with the same velocities as the original stops and their momenta tend to cancel each other out, leaving little missing transverse energy (MET) in the signal. Consequently, it is difficult to be distinguished from the SM top pair production background and it is still possible to have a relatively light $\tilde{t}_1$. There have been studies trying to identify useful variables to probe this compressed region but the reach is limited~\cite{Alves:2012ft,Han:2012fw,Kilic:2012kw,Dutta:2013gga}. A more promising strategy is to consider the stop pair production with a hard initial state radiation (ISR) jet, then the neutralinos are boosted in the opposite direction to the ISR jet, giving rise to some MET. It may have a discovery reach up to $m_{\tilde{t}_1}\sim 400-500$~GeV at LHC 13 TeV with 300 fb$^{-1}$~\cite{Hagiwara:2013tva,An:2015uwa,Macaluso:2015wja,Cheng:2016mcw}.

Since naturalness needs both stops to be not too heavy, if $\tilde{t}_1$ happens to lie in the compressed region, it may be easier to search for $\tilde{t}_2$ even though it is heavier. Indeed, 
both ATLAS and CMS have performed searches for the heavier stop for $m_{\tilde{t}_1} \approx m_t + m_{\tilde{\chi}_1^0}$ so that $\tilde{t}_1$ escapes the detection~\cite{Aad:2015pfx,Aad:2014mha,Khachatryan:2014doa,ATLAS:2016tpc}.  These searches adopted the simplified model approach, assuming that the heavier stop decays to the lighter stop plus a $Z$ or a Higgs boson ($ \tilde{t}_2 \to \tilde{t}_1 + Z/h$) with a 100\% branching fraction. The exclusion limit for the $\tilde{t}_2$ mass goes up to $\sim 730$ GeV for the $\tilde{t}_2 \to Z \tilde{t}_1$ decay mode with 13 TeV Run 2~\cite{ATLAS:2016tpc} and about 600 GeV for the $\tilde{t}_2 \to h \tilde{t}_1$ decay mode with the 8 TeV Run 1 data~\cite{Aad:2015pfx} (the corresponding Run 2 analysis has not appeared yet). 

The simplified model approach is a good strategy if there is a dominant decay channel or the search limit is dictated by a certain decay channel. One can easily recast the search result to a wide range of models which have similar decay processes and final states, by rescaling the cross sections and branching ratios.
However, it is seldom a good approximation for the system of the two stops. Because the left-handed stop and sbottom belong to an $SU(2)_W$ doublet, there is always a sbottom with mass within the vicinity of the two stops. The presence of the sbottom will give additional decay modes of $\tilde{t}_2$. In addition, there could be other charginos and neutralinos lighter than $\tilde{t}_2$ besides $\tilde{\chi}^0$. If this is the case, there will also be decay chains going through them as intermediate states. As a result, the $\tilde{t}_2$ decays often have many decay channels without a dominant one~\cite{Eckel:2014wta,Guo:2013iij}. Different decay channels have different final states and hence require different signal selection criteria. It makes the $\tilde{t}_2$ search in the compressed region in a realistic scenario more complicated than simply rescaling the simplified model analysis.

An alternative approach to the simplified model is the pMSSM~\cite{Djouadi:2002ze,Berger:2008cq,CahillRowley:2012cb}, which parametrizes MSSM with some modest assumptions. The assumptions include $R$ parity conservation with the lightest neutralino being the LSP, minimal flavor violation at the TeV scale with no CP violation in the SUSY sector, and degenerate sfermion masses for the first two generations. It contains 19 phenomenological parameters and a scan over these parameters generates a large model samples for phenomenological studies.
However, if one is interested in the stop system, scanning over the full 19-parameter space may be an overkill.\footnote{A light stop study in the pMSSM approach can be found in Ref.~\cite{Belanger:2015vwa}.} To study the possible interesting decay patterns of the stop and sbottom system and their experimental signals one should focus on the most relevant parameters. This is the approach taken in this paper. We  divide the models into two scenarios. In the first scenario all neutralinos and charginos except the LSP are heavier than $\tilde{t}_2$ so they decouple. The only relevant particles are $\tilde{t}_1$, $\tilde{t}_2$, $\tilde{b}_1$, and $\tilde{\chi}_1^0$, whose masses and interactions are only governed by a few parameters. We scan through them and find model points with different characteristic decay patterns. We identify categories of signal channels which may be sensitive to various final states of the decay chains and study the signal significance over the backgrounds. The real search reach may require division and combination of many different channels. In the second scenario we consider additional charginos and neutralinos below the mass of $\tilde{t}_2$, which can introduce even more complicated decay patterns. We perform the similar study as in the first scenario for the model points where the additional charginos and neutralinos play important roles in the decay chains.  

This paper is organized as follows. In Sec.~\ref{sec:MSSM}, we discuss the spectrum of the third generation squarks in MSSM based on the naturalness and the  Higgs boson mass. We also summarize the current experimental constraints. In Sec.~\ref{sec:benchmark}, we consider some benchmark points for the stop and sbottom spectrum where $\tone$ is hidden in the compressed region. The benchmark models are divided into two types, depending on whether there are additional neutralinos and charginos which can appear in the decay chains of $\ttwo$ and $\bone$, We present the branching ratios of various decay chains and the fractions of final states of these benchmark points. In Sec.~\ref{sec:collider}, we discuss categories of signals for the $\ttwo$ and $\bone$ searches when $\tone$ is hidden. We perform simplified collider studies for the benchmark models to explore the future sensitivities at the 14 TeV LHC. The conclusions are drawn in Sec.~\ref{sec:conclusions}.
The compatibilities of our benchmark points with the current experimental constraints are examined in the Appendix.

\section{Stop Masses in MSSM}
\label{sec:MSSM}

\subsection{Argument of Naturalness}
In MSSM, after minimizing the Higgs potential, the $Z$ boson mass is given by~\cite{Martin:1997ns}
\begin{equation}
\frac{m_z^2}{2}=\frac{m_{H_d}^2+\delta_{H_d}^2-\tan\beta^2(m_{H_u}^2+\delta_{H_u}^2)}{\tan\beta^2-1}-\mu^2,
\label{eq:mz}
\end{equation}
where $m_{H_u}^2$ and $m_{H_d}^2$ are the soft SUSY breaking masses of the $H_u$ and $H_d$ doublets at an UV cutoff scale $\Lambda$, $\delta_{H_u}$ and $\delta_{H_d}$ represent the radiative contributions to the soft SUSY breaking masses below the cutoff, and $\mu$ is the SUSY-preserving Higgs mass parameter which is also the approximate Higgsino mass. To avoid fine-tuning on the $Z$ mass, there should not be a large cancellation among various terms in the above equation. In particular, the radiative correction $\delta_{H_u}$ receives the largest contribution from the stop loops:
\begin{equation}
\delta_{H_u}^2 = - \frac{3y_t^2}{8\pi^2}(m_{Q_3}^2+m_{u_3}^2+|A_t|^2) \ln \left(\frac{\Lambda}{\mu_{\text{IR}}}\right)
\end{equation}
where $\mq, \muu$ are soft-breaking mass terms for the left-handed top-bottom doublet and the right-handed top squarks, $A_t$ is the trilinear soft-SUSY breaking of the corresponding Yukawa interaction, and $\mu_{\text{IR}}$ is taken to be the geometric average of the stop masses $m_{\tilde{t}} = \sqrt{m_{\tilde{t}_1} m_{\tilde{t}_2}}$. The tuning of $m_Z^2$ due to the stop mass contribution
$
\frac{m_Z^2/2}{|\delta_{H_u}^2|}
$
is already at the level of $\lesssim 1\%$ for $\mq, \muu \sim 1$~TeV and $\Lambda \sim 100$ TeV~\cite{Harnik:2003rs}. Therefore, naturalness argument would prefer both stops to have masses below or around 1 TeV.

\subsection{Higgs boson mass}

At the tree level, MSSM also predicts the light Higgs boson mass to be less than the $Z$ mass:
\begin{equation}
m_{h,\text{tree}}^2 =m_z^2\cos^22\beta .
\end{equation}
This contradicts the observed Higgs mass at 125 GeV. The loop corrections can raise the Higgs boson mass to evade the tree-level upper bound of $m_Z$. However, to reach 125 GeV the loop contributions must be significant. The dominant contribution comes from the stop loop, which implies constraints on the masses of the stop sector. The Higgs boson mass including the leading one-loop stop contribution is given by~\cite{Okada:1990gg,Casas:1994us,Carena:1995bx}
\begin{equation}
m_h^2=m_z^2\cos^22\beta+\frac{3m_t^4}{4\pi^2v^2} \left[ \log\frac{m_{\tilde{t}}^2}{m_t^2} +\frac{X_t^2}{m_{\tilde{t}}^2}
\left(1-\frac{X_t^2}{m_{\tilde{t}}^2}\right)
\right] ,
\end{equation}
where $\mtt = \sqrt{\mtone \mttwo}$ as defined previously and $X_t = A_t - \mu \cot \beta$ is the stop mass mixing parameter. From the formula one can see that without the $X_t$ term, the stop masses need to be raised to very high values in order to generate a Higgs massif 125 GeV. This would be in severe conflict with naturalness. To minimize the fine-tuning, the second term in the bracket should be large and the one-loop correction is maximized when $X_t = \sqrt{6}m_{\tilde{t}}$. Such a large $X_t$ implies a large mixing between the left-handed and right-handed stops, which has interesting phenomenological implications.\footnote{If there are additional contributions to the Higgs quartic coupling, such as in the Next-to-Minimal Supersymmetric Standard Model (NMSSM)~\cite{Ellwanger:2009dp}, it is easier to have a 125 GeV Higgs boson with light stops compatible with naturalness~\cite{Hall:2011aa,Beuria:2015mta}. The stop mixing does not need to be large in that case and the stop spectrum could be more compact. Nevertheless, the results in this paper also applies to a large region of parameter space in NMSSM.}

\subsection{Stop and sbottom masses}
The mass matrices for the stops, $\tt_L, \tt_R$, and sbottoms, $\bt_L\,$ and $\bt_R$ are given by
\begin{eqnarray}
M_{\tilde{t}}^2 =\left(\begin{array}{cc}
\mq^2 + m_t^2 +\Delta_{\tilde{u}L} & m_t X_t \\
m_t X_t & \muu^2 + m_t^2 +\Delta_{\tilde{u}R}
\end{array} \right),
\end{eqnarray}
\begin{eqnarray}
M_{\tilde{b}}^2 =\left(\begin{array}{cc}
\mq^2 + m_b^2 +\Delta_{\tilde{d}L} & m_b X_b \\
m_b X_b & \md^2 + m_b^2 +\Delta_{\tilde{d}R}
\end{array} \right)
\end{eqnarray}
where $X_b = A_b - \mu \tan\beta$ is the term related to $\bt_L$ and $\bt_R$ mixing from trilinear couplings, and 
\begin{equation}
\Delta_{\tilde{u}L}= (\frac{1}{2}-\frac{2}{3} \sin^2\theta_W) \cos 2\beta\, m_Z^2, \quad \Delta_{\tilde{u}R}=-\frac{2}{3}\sin^2\theta_W \cos 2\beta\ m_Z^2,
\end{equation}
\begin{equation}
\Delta_{\tilde{d}L}= (-\frac{1}{2}+\frac{1}{3} \sin^2\theta_W) \cos 2\beta m_Z^2, \quad \Delta_{\tilde{d}R}=\frac{1}{3}\sin^2\theta_W \cos 2\beta\ m_Z^2
\end{equation}
represent the $D$-term contributions.

Notice that in the limit where there is no mixing, both $\tt_L$ and $\bt_L$ masses are controlled by the soft breaking mass $\mq$ and they are expected to be nearly degenerate if $\mq \gg m_t,\, m_Z$, with $m_{\tt_L} \gtrsim m_{\bt_L}$. However, with a large $X_t$, there is a significant mixing between $\tt_L$ and $\tt_R$ and the mass spectrum of the stops will be modified. The two mass eigenstates are repelled from one another by the mixing term and the mass gap between them further increases. Consequently, there is at least one sbottom lighter than the heavier stop. The spectrum of the stop and sbottom sector has important implications for their decay patterns and collider searches as we will see.


To scan the MSSM parameter space we specify the parameters at the cutoff scale $\Lambda$ which is taken to be 100 TeV. Since we are focusing on the spectrum of the third generation squark (and neutralino/chargino for the decay patterns), we decouple the first two generation of sfermions and third generation sleptons by setting their soft SUSY-breaking masses to 3 TeV. We also set the gluino mass to 2.5 TeV, beyond the current and near future reaches.\footnote{A lighter gluino does not affect the direct stop and sbottom productions. However, it may give the first SUSY signals at the LHC. Its decays through stops and sbottoms will mix into the signals for direct stop and sbottom productions, so we choose a heavier gluino to avoid this complication.}
In order to generate stop masses which are potentially within the LHC Run 2 reach, the diagonal stop soft breaking masses $\mq$ and $\muu$ are varied from 250 GeV to 1.4 TeV. The $X_t$ term is scanned from $-3m_{\tilde{t}}$ to $3m_{\tilde{t}}$, where $\mtt=\sqrt{\mq  \muu} $ as mentioned before. On the other hand,  the right-handed sbottom soft-breaking mass $\md$ is varied from 100 to 3000 GeV. For the Higgs sector, $\tan\beta$ varies from 2 to 50 and the Higgsino mass parameter $\mu$ varies from 100 GeV to 3 TeV. Although the $SU(2)$ and $U(1)$ gaugino masses $M_2$ and $M_1$ have little effects in the stop/sbottom spectrum or the Higgs boson mass, the presence of the neutralinos and charginos can affect the decay chains of the stop/sbottom. Thus, we let $M_1$ and $M_2$ vary in the ranges of 50--1500 GeV and 250-1500 GeV respectively.

\begin{figure}[t]
\captionsetup{justification=raggedright,
singlelinecheck=false}
\begin{center}
\begin{subfigure}[b]{0.8\textwidth}
\includegraphics[width=\textwidth]{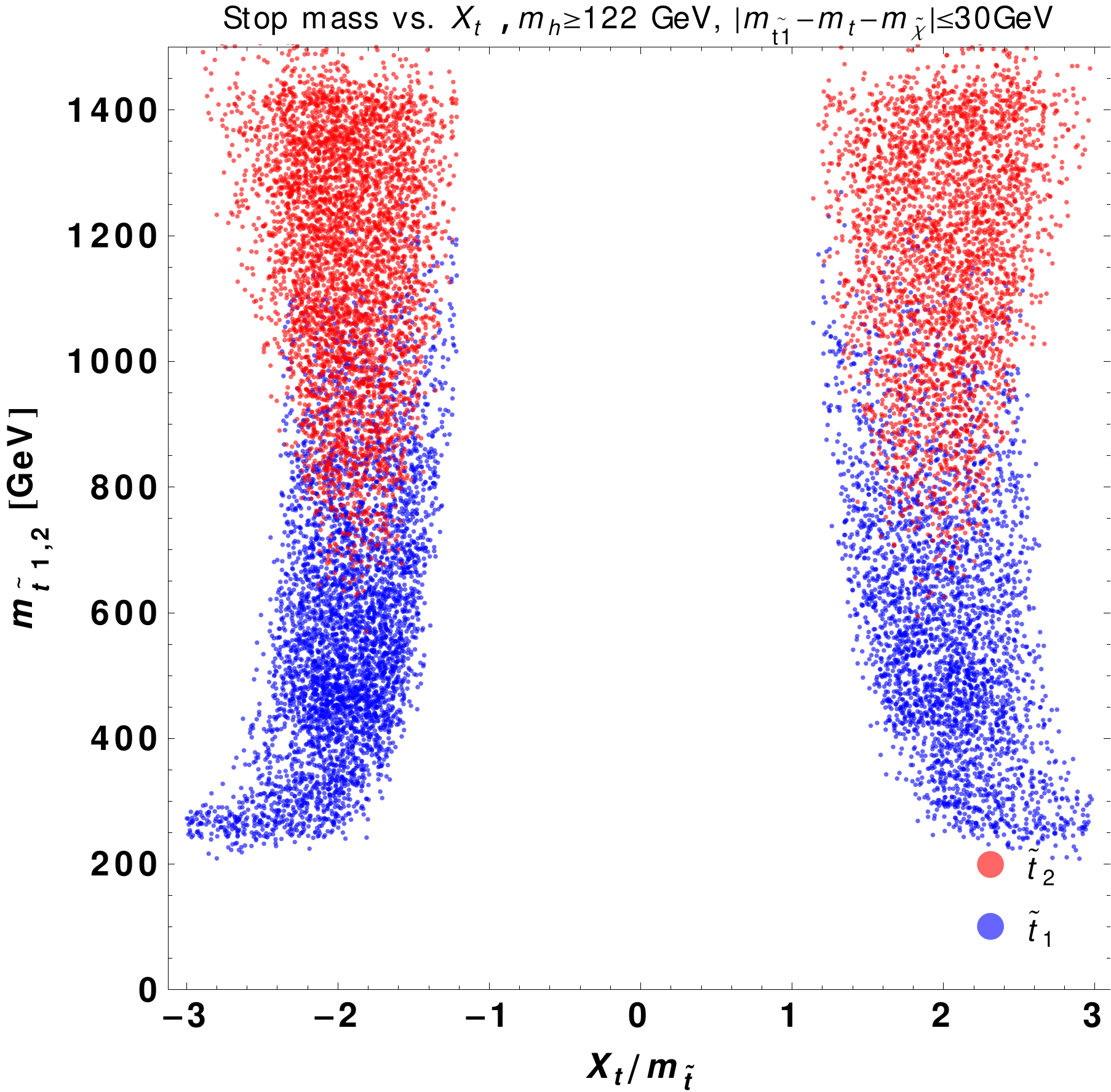}

\end{subfigure}
\caption{The model points that satisfies $m_h>122$GeV presented in $X_t  $ vs. $\mtt $ plane. Each red dot represents the second stop mass in GeV, and the blue one represents the lightest stop. All sample points presented have a proper LSP with a mass $|\mtone - \mcone - m_t| \leqslant30$ GeV.\label{fig:stop_vs_Xt}
  }
\end{center}
\end{figure}

We use FeynHiggs~\cite{Feynhiggs} to generate the SUSY spectrum and to calculate the Higgs boson mass. Given the uncertainties in different approaches in the Higgs mass calculation and higher order corrections, we require the resulting Higgs boson mass to be bigger than 122 GeV as our selection criterion.\footnote{The effective field theory approach generally gives a lower Higgs boson mass~\cite{Vega:2015fna} so we do not impose an upper limit on the Higgs mass.} The masses of the two stop mass eigenstates and the corresponding mixing term $X_t$ which can satisfy the Higgs boson mass requirement are shown in Fig.~\ref{fig:stop_vs_Xt}

\begin{figure}[t]
\captionsetup{justification=raggedright,
singlelinecheck=false}
\begin{center}
\begin{subfigure}[b]{0.6\textwidth}
\includegraphics[width=\textwidth]{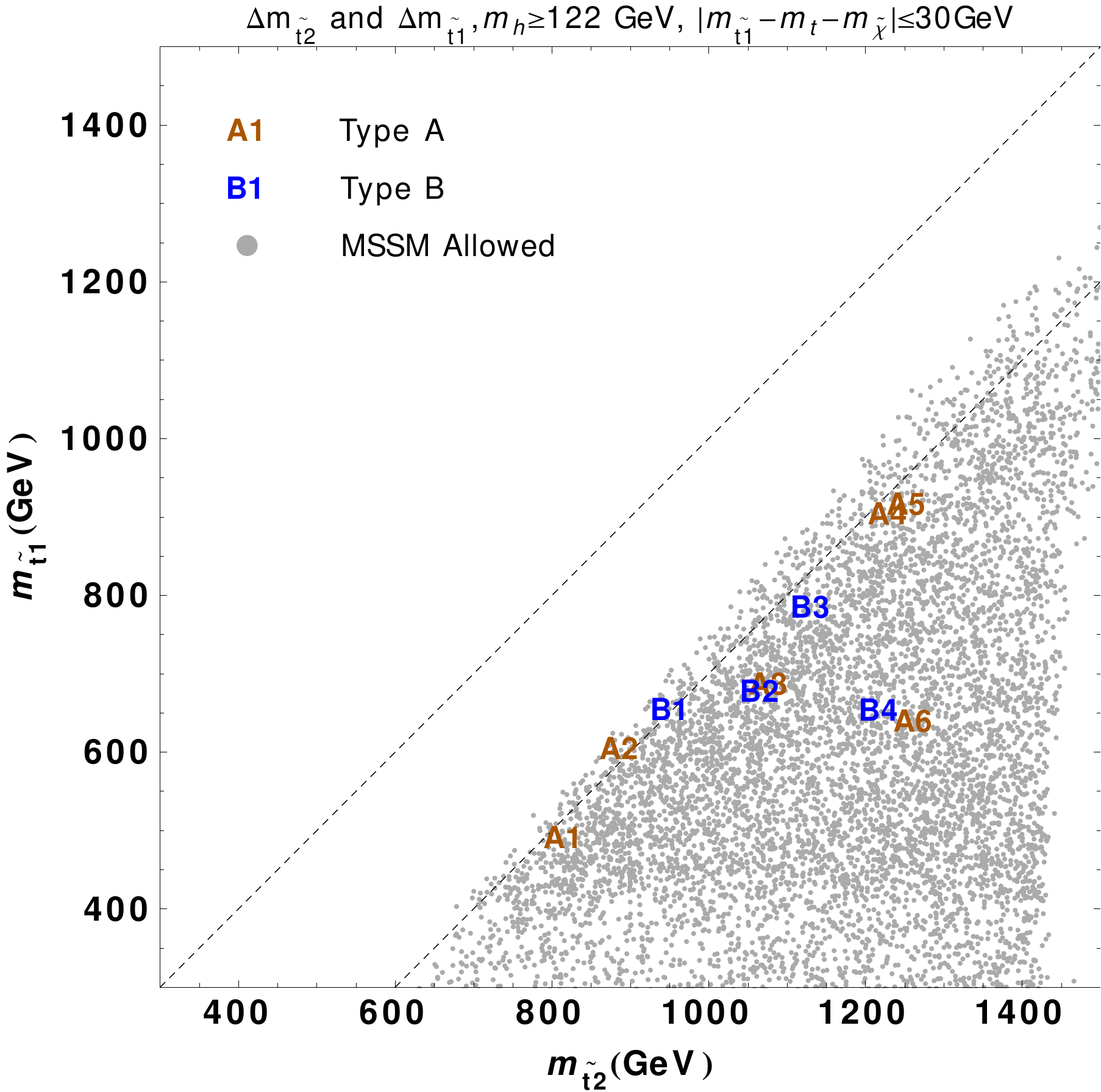}

\end{subfigure}
\caption{The mass and mass difference between the second and first stops. Two diagonal dashed lines represent $m_{\tilde{t}2}-m_{\tilde{t}1}=0,300$ GeV. Colored points are the benchmark models.
\label{fig:stop1_stop2}}
\end{center}
\end{figure}

\begin{figure}[t]
\captionsetup{justification=raggedright,
singlelinecheck=false}
\includegraphics[scale=0.6]{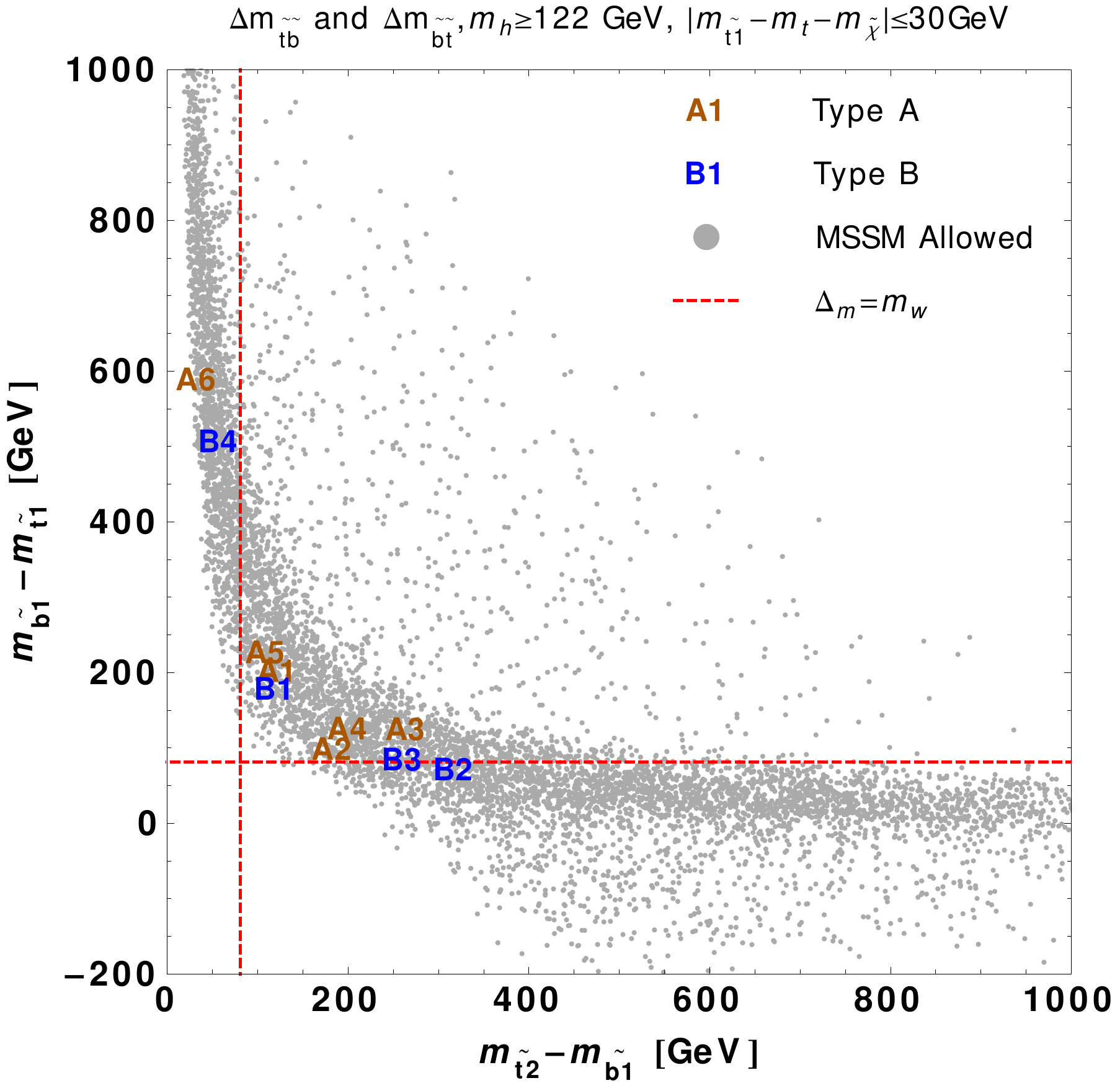}
\caption{A detailed look at $m_{\tilde{t}_2}-m_{\tilde{b}_1}$ vs. $m_{\tilde{b}_1}-m_{\tilde{t}_1}$ plane, including the right hierarchy only. The vertical dashed line represents the mass relation $m_{\tilde{t}_2}-m_{\tilde{b}_1} = m_W$. The on-shell charged current decay is kinematically forbidden to the left of this line. The horizontal lines are $m_{\tilde{b}_1}-m_{\tilde{t}_1}= m_W$. Colored points are the benchmark models.\label{fig:stop_sbottom}}
\end{figure}

We can see that the lightest stop mass can be as low as 250 GeV in extreme cases, while the second stop can be as low as around 600 GeV. 
When it comes to how the sbottom and the second stop decay, an interesting question is the mass differences between the stops and the sbottom. In Fig.~\ref{fig:stop1_stop2} we show $\mtone$ vs. $\mttwo$ for allowed points. There is always a significant gap between the two stop masses due to the large mass mixing term. Most points have a mass difference greater than 300 GeV, which means that $\ttwo \to \tone + Z(h)$ decays are always open. Fig.~\ref{fig:stop_sbottom} shows the mass difference between $\bone$ and $\tone$ vs. the mass difference between $\ttwo$ and $\bone$. We see that $\bone$ is always lighter than $\ttwo$. There are also points with $\bone$ lighter than $\tone$ but for these points the $\bone$ search will provide the strongest constraint~\cite{AdeelAjaib:2011ec,Alvarez:2012wf,Lee:2012sy,Bi:2012jv,Dutta:2015hra,Han:2015tua,Beuria:2016mur,CMS:2016xva}. For all points we have either $\mbone -\mtone > m_W$ or $\mttwo -\mbone > m_W$, and there is also a significant fraction of points where both inequalities hold. For $\mttwo -\mbone > m_W$, the decay channel $\ttwo \to \bone +W$ is open, which has not been considered in current $\ttwo$ searches. Similarly $\bone \to \tone +W$ will be open if $\mbone -\mtone > m_W$~\cite{Li:2010zv,Datta:2011ef}. These decay channels should be included in searches for $\ttwo$ and $\bone$ since they occur naturally in MSSM. There will be even more possible decay channels if some charginos and neutralinos have masses between these stop and sbottom states. We will perform some benchmark point studies in the rest of the paper to point out the final states and channels that are relevant for the second stop and sbottom searches.

\subsection{Current experimental bounds}

The third generation squarks are extensively searched at the LHC experiments. Here we give a brief summary of the experimental constraints most relevant to our discussion. 

For $\tone$ decaying 100\% to $t \tilde{\chi}_1^0$, 13 TeV Run 2 results based on the integrated luminosity $\sim 13$ fb$^{-1}$ exclude $\tone$ mass up to 860 GeV for a light ($\lesssim 250$ GeV) $\cone^0$~\cite{CMS:2016hxa,CMS:2016vew,ATLAS:2016jaa,ATLAS:2016ljb}.
However, this does not apply to the compressed region when the mass difference between $\tone$ and $\cone^0$ is small. For $\mtone -m_{\tilde{\chi}_1^0} \approx m_t$, while there was no bound from Run 1, with Run 2 data ATLAS has obtained a limit on $\mtone$ of 380 GeV~\cite{ATLAS:2016jaa} using the variable proposed in Ref.~\cite{An:2015uwa}. For even smaller mass difference, there are some constraints from several search modes $\tone \to Wb \cone^0$, $c\cone^0$, or $b f f' \cone^0$ depending on the mass difference~\cite{Hikasa:1987db,Muhlleitner:2011ww,Boehm:1999tr,Das:2001kd}. The bounds on $\mtone$ in these cases are around 300 GeV from the Run 1 data~\cite{Aad:2015pfx,Khachatryan:2016pup,Aad:2014kra,Aad:2014nra,Chatrchyan:2013xna,Khachatryan:2015pot,Khachatryan:2015wza,Khachatryan:2016pxa}. The most recent Run 2 analyses exclude a stop mass up to 365 GeV for $\mtone -m_{\tilde{\chi}_1^0}= 90$~GeV on the 3-body decay mode $\tone \to Wb \cone^0$~\cite{ATLAS:2016xcm}, and $\sim 450 (360)$ GeV for the 4-body decays $b f f' \cone^0$ in the fully hadronic final state (opposite-sign leptons) search~\cite{CMS:2016hxa,CMS:2016zvj}. For $\bone$ decaying 100\% to $b+\cone^0$, The 13 fb$^{-1}$ Run 2 results exclude $\mbone $ up to $\sim 1$ TeV for $m_{\cone^0}$ up to $\sim 400$ GeV~\cite{CMS:2016xva}.
For $\bone$ decaying to $t+\cone^-$ and then $\cone^- \to W^- + \cone^0$, the 13 fb$^{-1}$ Run 2 analysis reached a mass limit of 690 GeV for a light $\cone^0$, assuming $m_{\cone^\pm} = m_{\cone^0} +100$ GeV, while $m_{\cone^0} \lesssim 260$ GeV are also excluded for $m_{\bone} \approx 540$ GeV~\cite{ATLAS:2016kjm}.

If the LSP is Wino-like or Higgsino-like, then one expects that there is a chargino state with its mass close to that of the LSP. The decay $\tone \to b \cone^\pm$ will be open as long as $\tone$ is heavier. The decay products of the $\cone^\pm$ to $\cone^0$ are likely to be too soft to be detected. The signal is similar to the $\bone$ search discussed above with the similar limit.

For $\mtone \approx m_t + m_{\tilde{\chi}_1^0}$, the current bound on the $\tone$ mass is rather weak. Both ATLAS and CMS performed searches for the $\ttwo$ states with the decays to $Z \tone $, $h \tone $. The bound on $\mttwo$ from the most recent ATLAS Run 2 analysis is up to 730 GeV for 100\% decay to $Z \tone$~\cite{ATLAS:2016tpc}. The analysis for the $h \tone$ mode with Run 2 data has not come out yet. The limit from Run 1 data is about 600 GeV~\cite{Aad:2015pfx,Khachatryan:2014doa} and similar limits were obtained if $\ttwo$ decays to a mixture of the $Z \tone$, $h \tone $, and $t\cone^0$ final states.

\section{Benchmark Points for Case Studies}
\label{sec:benchmark}

We are interested in searches for the heavier stop and sbottom in the case when the lightest stop is hidden. Therefore we choose the benchmark points of our study to satisfy $0\leqslant\mtone-m_t-\mcone \leqslant20$ GeV where the bound is weakest. Of course, for a heavier $\tone$ the difference can be bigger from current bounds. For $\mtone$ being much closer to $\mcone^0$, there have been several studies focusing on these scenarios~\cite{Beuria:2015mta,Kaufman:2015nda,Ghosh:2013qga}. We also assume that $\cone^0$ is Bino-like so that there is no nearby chargino state. Otherwise the $\tone \to b \cone^\pm$ search would provide a strong constraint. Moreover, if $\bone$ is close to or even lighter than $\tone$, then the sbottom search in $\bone \to b\cone^0$ decay will provide the strongest constraint. As we see from the summary in the last section, the constraint on $b$-jets + MET is very strong and the exclusion limit has reached $\sim 1$ TeV for these decay modes. Thus, we will focus on points where $\bone$ is somewhat heavier where other decay modes such as $\bone \to W \tone$ are open, so that the traditional search based on $\bone \to b\cone^0$ is not effective. We assume that the gluino is heavy enough so that the stops and sbottoms are dominated by the direct pair production. Otherwise the gluino cascade decays would be the strongest probe. 

Given the spectrum $\mcone^0 < \mtone < \mbone < \mttwo$, we can find two main classes of model points depending on whether there are other neutralinos and charginos lying between them. We denote Type A models to have the lightest charginos and the second lightest neutralinos heavier than our second stop, so that they decouple from our discussion, given that their direct pair production rate is much smaller than that of squarks of the same mass.
In Type B models, the second neutralino and accompanying charginos are lighter than $\ttwo$ so that they may appear in the cascade decays of $\ttwo$. These neutralinos and charginos may be either Wino-like or Higgsino-like, or both. From the naturalness point of view, it is preferable to have the Higgsinos not too heavy. In addition, Winos couple to the squarks via the $SU(2)$ gauge coupling instead of the large top Yukawa coupling. The branching fractions of stop and sbootom decays through the Winos are often small. Then the decay patterns are mostly similar to the Type A models. Therefore for Type B models we focus on the cases where the relevant charginos and neutralinos are mostly Higgsino-like. The benchmark points are selected to be compatible with the current experimental constraints as examined in the Appendix.

\subsection{Type A}
From the parameter space scan described earlier, we select several benchmark points of Type A spectrum, listed in Table~\ref{table:A_spectrum}.\footnote{In the first version of the paper, the A1 point has been ruled out after the release of new data at ICHEP 2016, so it is removed and the original A2 is moved to A1. We add two new benchmark points A2 and A3 which have similar decay patterns as the original A1, but with heavier spectra. The original A3-A5 are shifted to A4-A6.} The range of $\ttwo$ and $\bone$ masses is chosen from $\sim 700$ GeV up to $\sim 1.2$ TeV. A1 and A2 have relatively light spectra which are not far from the current bounds. They may soon be discovered or excluded at LHC Run 2 with more luminosities. Benchmarks A4--A6 have heavy $\ttwo$ and $\bone$ which are close to the reach limits of the 14 TeV LHC. 
As expected, to obtain a Higgs boson mass close to 125 GeV, these benchmark points all have large mixing $|X_t/\mtt|$ between the left- and right-handed stops.
\begin{table}[ht]
\begin{center}
\begin{tabular}{|c|c|c|c|c|c|c|}
\hline 
Spectrum & A1 & A2 & A3 & A4 & A5 & A6 \\ 
\hline \hline 
$m_{\tilde{t}_2}$(GeV) & 815.4 & 887.1 & 1077.3 & 1230.6 & 1253.2 & 1262 \\ 
\hline 
$m_{\tilde{b}_1}$(GeV) & 693.0 & 704.5 & 812.8 & 1029.5 & 1143.8 & 1229\\ 
\hline 
$m_{\tilde{t}_1}$(GeV) & 491.0 & 605.5 & 687.6 & 904.0 & 916.5 & 640.1\\ 
\hline 
$m_{\tilde{\chi}_1^0}$(GeV) & 304.9 & 414.2 & 498.0 & 710.8 & 724.2 & 459.4\\ 
\hline 
$X_t/m_{\tilde{t}}$ & -1.81 & 1.58 & -2.17 & -1.84 & -1.82 & 1.51\\
\hline 
$m_h$(GeV) & 122.8 & 122.7 & 123.4 & 124.9 & 124.6 & 122.3\\  
\hline
\end{tabular} 
\end{center}
\caption{Spectra of Type A Benchmark points.\label{table:A_spectrum}}
\end{table}

Due to the large mixing term, it is typical to have a large mass gap between $\tone$ and $\ttwo$, leaving a relatively large phase space for $\ttwo \to \tone + Z/h$ decays. In the scanned parameter space it is also common to have the mass difference $\mttwo - \mbone \geq m_W$, which opens up the decay mode $\ttwo \to \bone + W$. This decay mode has not been considered by the experimental analysis of $\ttwo$ searches.
In Table~\ref{table:A_decay} we list the branching ratios of various decay modes of $\ttwo$ and $\bone$ for the benchmark points. The branching ratios are calculated by SDECAY~\cite{Muhlleitner:2003vg}.
\begin{table}[ht]
\captionsetup{justification=raggedright,
singlelinecheck=false}
\begin{center}
\begin{tabular}{|c|c|c|c|c|c|c|c|c|c|c|}
\hline 
Channel & A1 & A2 & A3 & A4 & A5 & A6\\
\hline \hline
$\tilde{t}_2 \to \tilde{b}_1 + W^+$ & 16.5 & 42.0 & 48.2 & 42.1 & 8.1 & 0 \\   
\hline 
$\tilde{t}_2 \to \tilde{t}_1 + Z$ & 74.5 & 47.6 & 44.6 & 52.9 & 79.2 & 53.1 \\ 
\hline 
$\tilde{t}_2 \to \tilde{t}_1 + h$ & 5.9 & 3.9 & 5.0 & 2.5 & 10.7 & 45.6 \\
\hline 
$\tilde{t}_2 \to t+\tilde{\chi}_1^0$ & 3.1 & 6.5 & 2.2 & 2.5 & 2.0 & 1.2\\
\hline \hline
$\tilde{b}_1 \to \tilde{t}_1 + W^-$ & 99 & 90.1 & 98.0 & 98.3 & 99.5 & 99.3 \\ 
\hline 
$\tilde{b}_1 \to b+\tilde{\chi}_1^0$ & 1.0 & 9.9 & 2.0 & 1.7 & 0.5 & 0.7\\

\hline 
\end{tabular}
\end{center}
\caption{Branching ratios of the major decays of $\ttwo$ and $\bone$ for the Type A benchmark models. The rest of the decay branching ratio is $\ttwo$ and $\bone$ decaying to the corresponding quarks and an LSP.\label{table:A_decay}}
\end{table}

As it can be seen, in Type A models, the $\tilde{t}_2 \to \tilde{t}_1 Z$ decay branching fraction is always significant, and it is dominant for A1 and A5. The $\tilde{t}_2 \to \tilde{t}_1 h$ branching ratio is smaller than the $\tone Z$ branching ratio. It is controlled by the difference between two diagonal soft breaking masses. More specifically, the $h\ttwo \tone$ coupling is proportional to $\cos 2 \theta_t$ where $\theta_t$ is the mixing angle diagonalizing the stop mass matrix. If the two diagonal soft breaking masses are exactly equal, $\muu = \mq$, then $\theta_t=\frac{\pi}{4}$ and the mass eigenstates $\tone,\ttwo$ will be equal mixtures between $\tilde{t}_L$ and $\tilde{t}_R$. In this case the $h\ttwo \tone$ coupling cancels between the contributions coming from the left-handed and the right-handed stops, and the $\ttwo \to \tone + h$ decay will be highly suppressed.
In models where $\ttwo \to \bone + W$ decay is kinematically allowed, the branching ratio of this channel increases rapidly as the allowed phase space expands, and easily becomes comparable to the $\tone + Z$ decay channel, as in models A2--A4. The model A6 is chosen such that the $\ttwo \to \bone + W$ decay is closed while the $\tilde{t}_2 \to \tilde{t}_1 h$ branching ratio is significant.

On the other hand, in Type A spectrum, $\bone$ decays predominately  
to $\tone W$ if the phase space allows. This is not covered by the current experimental searches for direct pair-production of $\bone$~\cite{Aad:2015pfx,Khachatryan:2015wza,Khachatryan:2016xvy,CMS:2014dpa,Chatrchyan:2013fea,Aad:2014pda,Khachatryan:2016kod}. We would advocate that this decay mode should be included in the future direct $\bone$ pair production search analysis. In these benchmark models,  the branching ratios of the direct decays of $\ttwo$ and $\bone$ to an LSP plus $t$ or $b$ are small, so the searches using these decays will not be effective.

In Table~\ref{table:A_final} we list the fractions of the final states in terms of $t, b, W, Z, h$ (without their subsequent decays) of the $\ttwo \ttwo$ and $\bone \bone$ pair productions, aside from a pair of $\cone^0$'s which are implicitly understood. The recent advances in jet substructure techniques to tag hadronically decayed top quarks may help to identify final states which contain them~\cite{Kaplan:2008ie,CMS:2009lxa}, so in the list we keep the top quark instead of its decay product in the final states. As a result, all $W$ bosons and $b$ jets listed in Table~\ref{table:A_final} are coming from SUSY particle decays instead of top decays.

\begin{table}
\captionsetup{justification=raggedright,
singlelinecheck=false}
\begin{center}

\begin{tabular}{|c|c|c|c|c|c|c|}
\hline 
 & A1 & A2 & A3 & A4 & A5 & A6\\ 
\hline \hline
$\sigma(\tilde{t}_2\tilde{t}_2) (fb)$ & 33.8 & 19.4 & 5.1 & 1.9 & 1.7 & 1.6\\ 
\hline \hline
$ttZZ$ & 55.5 & 22.6 & 19.9 & 28 & 62.7 & 28.2\\ 
\hline 
$ttZWW$ & 24.6 & 40.0 & 42.1 & 44.5 & 12.8 & 0\\ 
\hline 
$ttZh$ & 8.8 & 3.7 & 4.5 & 2.6 & 16.9 & 48.4\\ 
\hline 
$tt4W$ & 2.7 & 17.6 & 22.3 & 17.7 & 0.7 & 0\\ 
\hline 
$tthWW$ & 1.9 & 3.3 & 4.7 & 2.1 & 1.7 & 0\\ 
\hline 
$tthh$ & 0.4 & 0.1 & 0.3 & $\sim$0 & 1.1 & 20.8\\ 
\hline \hline
$\sigma(\tilde{b}_1\tilde{b}_1) (fb)$ & 94.5 & 85.2 & 34.5 & 7 & 3.3 & 1.9\\ 
\hline \hline
$ttWW$ & 98 & 81 & 96 & 96.6 & 99 & 98.6\\ 
\hline 
$tbW$ & 2 & 19.6 & 4 & 3.3 & 1.0 & 1.3\\  
\hline 

\end{tabular} 
\caption{14 TeV production cross sections and fractions of the final states (in terms of $t,\, b,\, Z,\, W,\, h$) of $\ttwo$ and $\bone$ for Type A models. All final states also contain an additional  $\tilde{\chi}^0_1\tilde{\chi}^0_1$ pair which becomes MET. In the list, the $W$ and $b$ jets in the final states are produced from squark decays rather than top decay.\label{table:A_final}}
\end{center}
\end{table}

It is pretty common for Type A models the heavy stop pairs produce an excess in $W$ bosons. Since each top quark itself gives another $W$ boson in its decay, the chances for the $\ttwo$ pairs in our model points to give a final state with 4 or more $W$'s varies up to $\sim 70 \%$. The high multiplicity of $W$ bosons can lead to excesses in same-sign dilepton and multiple lepton events. As a result, these types  of signals are also useful for $\ttwo$ searches in our benchmark models. Also, top pair associated with additional $ZZ$, $Zh$ or $hh$ constitutes considerable fractions of final states for some model points. They are the basis of the existing experimental searches for $\ttwo\ttwo$ direction production.
Sbottom pairs in our benchmark points predominately decay into $4W + 2b$ final state, which also leads to an excess in same-sign dilepton and multiple lepton signals.

\subsection{Type B}
In Type B models there are additional neutralinos and charginos lighter than $\ttwo$. To be distinct from the phenomenology of Type A models, we select benchmark points where these neutralinos and charginos have significant appearance in the decay chains from $\ttwo$. Because Higgsinos have larger couplings to the stops than the Winos, for the benchmark points we choose these neutralinos and charginos to be Higgsino-like. 

\begin{table}[ht]
\begin{center}
\begin{tabular}{|c|c|c|c|c|}
\hline 
Spectrum & B1 & B2 & B3 & B4  \\ 
\hline \hline 
$m_{\tilde{t}2}$(GeV) & 952.6 & 1067.1 & 1131.9 & 1265.4 \\ 
\hline 
$m_{\tilde{b}1}$(GeV)   &  832.5 & 749.1 & 870.4 & 1232.6 \\ 
\hline 
$m_{\tilde{t}1}$(GeV)  & 654.4 & 677.8 & 785.6 &  585.3 \\ 
\hline 
$m_{\tilde{\chi}_1^0}$(GeV)  & 478.6  & 499.5 & 594.8 &  407.7\\ 
\hline 
$m_{\tilde{\chi}_2^0}$(GeV)  & 774.4 & 702.8 & 824.1 &  823.3 \\ 
\hline 
$m_{\tilde{\chi}_3^0}$(GeV)  & 775.7 & 703.2 & 824.3 &  825.7 \\ 
\hline 
$m_{\tilde{\chi}^\pm}$(GeV)  & 772.3 & 699.5 & 820.7 &  822.1 \\ 
\hline 
$X_t/m_{\tilde{t}}$  & 1.65 & -1.78 & -1.74 &  -1.69 \\
\hline 
$m_h$(GeV)  & 124.0 & 124.7 & 124.9 &  123.5 \\  
\hline 
\end{tabular} 
\end{center}
\caption{Spectra of Type B benchmark points. $\tilde{\chi}_2^0, \tilde{\chi}_3^0, \tilde{\chi}^\pm$ are Higgsino-like.\label{table:B_spectrum}}
\end{table}
The spectra of Type B benchmark points are listed in Table~\ref{table:B_spectrum}.
The mass gap between $\ttwo,\bone$ and $\cc,\ctwo^0$ are chosen to be big enough to allow sufficient phase space for decays through the additional neutralinos or charginos, such as $\ttwo \to t \ctwo^0$ or $b\cc$. Therefore these benchmark points have relatively heavy $\ttwo$, and $\mttwo$ ranges from $\sim$ 900 GeV to 1.3 TeV. For a more compressed spectrum, the decays of $\ttwo$ and $\bone$ into second neutralino and charginos are suppressed by the phase space, then the decay patterns will be similar to those of Type A benchmark points. If $\cn^0_{2,3}, \cc$ are lighter than $\bone$, $\bone$ can also decay to $b\cn_{2,3}$ or $t\cc$ (if kinematically open), in addition the  $\ttwo \to b\cc, \, t\cn^0_{2,3}$ decays. The branching ratios of SUSY particle decays for the benchmark points are listed in Table~\ref{table:B_decay}.
\begin{table}[ht]
\captionsetup{justification=raggedright,
singlelinecheck=false}
\begin{center}
\begin{tabular}{|c|c|c|c|c|c|c|}
\hline 
Channel & B1 & B2 & B3 & B4\\
\hline \hline
$\tilde{t}_2 \to \tilde{t}_1 + Z$  & 58.2 & 24.4 & 30.2 &  30.2 \\
\hline 
$\tilde{t}_2 \to \tilde{b}_1+W^+$  & 12.9 & 36.6 & 38.0 &  0 \\  
\hline 
$\tilde{t}_2 \to \tilde{t}_1 + h$  & 3.3 & 9.6 & 7.8 &  25.8 \\  
\hline 
$\tilde{t}_2 \to t+\tilde{\chi}_2^0$  & 2.7 & 3.9 & 7.7 &  9.6 \\ 
\hline 
$\tilde{t}_2 \to t+\tilde{\chi}_3^0$  & 0.1 & 7.8 & 2.9 &  7.6 \\ 
\hline 
$\tilde{t}_2 \to b+\tilde{\chi}_1^\pm$  & 20.9 & 16.0 & 11.7 &  26.5 \\  
\hline \hline
$\tilde{b}_1 \to \tilde{t}_1 + W^-$  & 90.0 & 0 & 31.3 & 58.6 \\
\hline
$\tilde{b}_1 \to b + \tilde{\chi}_2^0$  & 3.7 & 35.1 & 19.7 &  12.6 \\ 
\hline
$\tilde{b}_1 \to b + \tilde{\chi}_3^0$  & 3.6 & 33.4 & 20.0 &  12.5\\ 
\hline
$\tilde{b}_1 \to t + \tilde{\chi}^\pm$  & 0 & 0 & 0 &  15.6 \\
\hline \hline
$\tilde{\chi}_2^0 \to t+ \tilde{t}_1$  & 0 & 0 & 0 &  92.1 \\
\hline
$\tilde{\chi}_2^0 \to h+ \tilde{\chi}_1^0$  & 94.2 & 1.8 & 96.9 &  7.1 \\
\hline
$\tilde{\chi}_2^0 \to Z+ \tilde{\chi}_1^0$  & 5.8 & 98.2 & 3.1 &  0.8 \\
\hline \hline
$\tilde{\chi}_3^0 \to t+ \tilde{t}_1$  & 0 & 0 & 0 &  86.9 \\
\hline
$\tilde{\chi}_3^0 \to h+ \tilde{\chi}_1^0$  & 4.4 & 96.1 & 1.7 &  1.2 \\
\hline
$\tilde{\chi}_3^0 \to Z+ \tilde{\chi}_1^0$  & 95.6 & 3.9 & 98.3 &  11.9 \\
\hline \hline
$\tilde{\chi}^\pm \to b+ \tilde{t}_1$  & 86.8 & 37.2 & 29.8 &  93.0 \\ 
\hline
$\tilde{\chi}^\pm \to W+ \tilde{\chi}_1^0$  & 13.2 & 62.8 & 70.2 &  7.0 \\ 

\hline 
\end{tabular}
\end{center}
\caption{Branching ratios of the major decays of $\ttwo$ and $\bone$ for the Type B benchmark models. The rest of the decay branching ratio is $\ttwo$ and $\bone$ decaying to the corresponding quarks and an LSP.\label{table:B_decay}}
\end{table}

From the Table~\ref{table:B_decay} we see that $\ttwo \to t \cn^0_{2,3}$ decay branching ratios are usually small, due to the phase space suppression. The $\ttwo \to b \cc$ branching ratio, on the other hand, can get up to about 1/4 for model B4. Similarly, the $\bone \to t \cc$ branching ratio is typically small but $\bone \to b \cn^0_{2,3}$ branching ratios can be quite big. In model B2, the sum of the branching ratios of $\bone \to b\cn^0_2$ and $b\cn^0_3$ is close to 70\%. The heavy neutralinos tends to decay to $\cone^0 + h/Z$, unless the decay channel to $t \tone$ is open (as in model B4), in which case it becomes the dominant decay channel. The chargino decays to $b\tone$ or $W\cone^0$. Which branching ratio is larger depends on the model point.

Similar to the Type A case, we can list the fractions of the final states of the $\ttwo\ttwo$ and $\bone\bone$ pair productions, in terms of $t, b, W, Z, h$ aside from the $\cone^0$. Because there are many more possible decay channels for Type B models, we only list the final states of a single decay chain in Table~\ref{table:B_final}. The complete final states can be obtained by simply squaring the Table. In addition to the final states which have been present in Type A models, Type B models can produce final states with a large number (up to 6) of $t$ or $b$ quarks. Therefore, search channels for multiple tops or bottoms could be interesting and important for some Type B spectra (e.g., B4).
\begin{table}
\captionsetup{justification=raggedright,
singlelinecheck=false}
\begin{center}

\begin{tabular}{|c|c|c|c|c|}
\hline 
 & B1 & B2 & B3 & B4\\ 
\hline \hline
$\sigma(\tilde{t}_2\tilde{t}_2) (fb)$  & 12.1 & 5.5 & 3.6 &  1.6 \\ 
\hline \hline
$tZ/bWZ$  & 58.5/0.5 & 28.5/13.1 & 33.3/7.7 &  31.2/0 \\ 
\hline 
$th/bWh$  & 5.8/0.5 & 17.2/12.0 & 15.3/7.4 &  26.6/0 \\ 
\hline 
$tbb$  & 18.1 & 6.0 & 3.5 &  24.6 \\ 
\hline 
$Wb$  & 2.8 & 21.5 & 19.2 &  1.9 \\ 
\hline 
$3t$  & 0  & 0 & 0 & 15.4 \\
\hline 
$tWW$  & 11.6 & 0 & 11.9 &  0 \\ 
 
\hline \hline
$\sigma(\tilde{b}_1\tilde{b}_1) (fb)$  & 29.6 & 58.1 & 22.1 &  1.9 \\
\hline \hline
$tW$  & 90 & 0 & 31.3 &  59.7 \\ 
\hline 
$ttb$  & 0 & 0 & 0 &  37.0 \\ 
\hline 
$hb$  & 3.7 & 32.7 & 19.4 &  1.0 \\ 
\hline 
$Zb$  & 3.6 & 35.8 & 20.3 &  1.6 \\

\hline 
\end{tabular} 
\caption{14 TeV production cross sections and fractions of the final states of $\ttwo$ and $\bone$ for Type B models. Only final states of a single $\ttwo$ or $\bone$ decays are listed. An additional LSP $\tilde{\chi}_1^0$ is implicitly understood. The fractions of the finals states of the squark pair can be easily obtained by squaring this table.\label{table:B_final}}

\end{center}
\end{table}

\section{Collider Studies for LHC 14 TeV}
\label{sec:collider}

Given the complex decay chains and many possible final state combinations of the second stop and the sbottom, we expect that there exist many possible signal channels. The best experimental reach may come from a combination of different signals. In this section we perform a rudimentary collider study for the benchmark points discussed in the previous section for the 14 TeV LHC and point out the interesting channels. Even though some of the channels were not considered in the $\ttwo$ simplified model analyses by ATLAS and CMS, most of them have been used in other SUSY or new physics searches. A purpose of this study is to point out their relevance for the $\ttwo$ and $\bone$ searches. These existing analyses could readily be adapted to the current case, maybe with some optimizations of cuts for the current scenario.  

For SUSY signals, we use MadGraph 5~\cite{Alwall:2014hca} to generate $\ttwo$ and $\bone$ pair events, with PYTHIA 6~\cite{Sjostrand:2006za} for parton showering and hadronization simulations. Detector simulations  were done by Delphes 3~\citep{deFavereau:2013fsa}, with the anti-$k_t$ jet algorithm~\cite{Cacciari:2008gp} and $\Delta R=0.5$. We use the CTEQ6L~\cite{Pumplin:2002vw} PDF in order to match the Snowmass  background simulations. All signal cross sections are normalized to NLO+NLL results with gluino decoupled~\cite{Borschensky:2014cia}.

We adopt the backgrounds generated by the Snowmass 2013 Energy Frontier Simulation group~\cite{Snowmass1,Snowmass2}. No-pileup effects are included for either the signals or the backgrounds. For this analysis the dominant background is the top-pair production, and top pair plus an extra boson ($W$, $Z$ or $h$). We also include other backgrounds such as the single/multiple boson(s) with jets, and the single top production.

\subsection{Basic cuts and tagging for backgrounds and signals}
\label{subsect:tagging}
The lepton and $b$ jet tagging efficiencies are taken to be the same as those of the Snowmass 2013 Energy Frontier Simulations. Additionally, we drop all leptons having a $\Delta R \leqslant 0.4$ with any jets. For both $e$ and $\mu$, we require them to have $p_T>15$~GeV and $\eta<2.5$. We will refer to these isolated leptons simply as leptons for the rest of our discussion. 

Each jet candidate is required to have $\eta<2.5$ and $p_T>25$~GeV, which is rather loose since our analysis is not very sensitive to non-$b$-tagged jets. The maximum tagging efficiency for $b$ jets is $\approx 70\%$. The $b$-jet mis-tagging rate from light flavors is $0.1\%$. A $b$ jet candidate must have $p_T>30$~GeV. Further selection rules for leptons and jets will be described in detail later.

Since all SUSY signals we discussed have two LSP which could lead to significant MET, an event with a higher MET is preferred, especially for those with no or only one lepton in the final state. On the other hand, SM background events with more leptons could get a large MET from the associated neutrinos. Therefore, for a higher lepton multiplicity we would have a lower the MET cut in order to include more signal events.  As a result, we set up a preliminary criteria based on MET: for all-hadronic or one-leptonic channels, all signal events are required to have a MET greater than 200 GeV. On the other hand, events with two/more than two leptons would be vetoed if their MET is less than 150/100~GeV. 

To further suppress the backgrounds, scalar $H_T$ could provide a good discrimination between the signal and backgrounds, since we are interested in the pair production of heavy particles. We require that each event should have scalar $H_T \geq 800$ GeV as a preliminary selection rule.

A pair of opposite-sign leptons of the same flavor can come from $Z$ decays. The $Z$ boson produced in $\ttwo$ decay could be somewhat boosted due to the large $\mttwo-\mtone$. To define the $Z$ candidate, we require that the opposite-sign same-flavor lepton pair  lepton pair to have $|\sqrt{(\Sigma p_\ell)^2} - m_Z | \leqslant10$~GeV and $\Delta R < 1.5$. 

Opposite-sign lepton pairs also often arise from SM background such as $t\bar{t}$. For our signals, $\ttwo$ pair decays usually produce extra $W/Z$ bosons. If they are boosted and decay hadronically, they may be tagged to help discriminating signals and backgrounds. We follow the method presented in Ref.~\cite{Han:2011ab}: A Cambridge-Archen jet algorithm~\cite{Dokshitzer:1997in} is adopted to identify fat jets from vector boson decays, with $\Delta R=0.8$ and $p_T>200$ GeV. Any fat jet constructed this way with a invariant mass between 60 and 110 GeV would be our vector jet candidate. Furthermore, we require $N$-subjettiness~\cite{Thaler:2010tr} $\tau_{21}=\frac{\tau_2}{\tau_1}< 0.5$, which means that its substructure is more likely to have two subjets rather than coming from QCD backgrounds. 

\subsection{Signal channels and benchmark results}

In brief, the SUSY signals we are looking for in this work can be understood as a bottom quark pair accompanied by multiple bosons ($W,Z$ or $h$) and a pair of missing neutralinos, with possible extra $t,b$ pairs.  Consequently, events with multiple $b$ jets are favored. Compared with the SM background such as $t\bar{t}$ events which can give two $b$ jets at parton level, our signals can have more $b$ jets in the final states. $\tone\tone hZ/\tone\tone ZZ/\tone\tone hh$ decays of $\ttwo\ttwo$ with an $h$ or $Z$ decay to a pair of $b$ quarks and $\tone\tone$ cascade to a pair of tops can give us 4 or more $b$ jets. Thus, requiring more than 3 $b$ jets in each event would be a good way to handle SM backgrounds~\cite{Aad:2014lra}. On the other hand, signal events with two or less $b$ jets would be overwhelmed by SM backgrounds such as $t\bar{t}$ production. Therefore, for events with $N_b \leqslant 2$ would need features such as a large multiplicity of leptons or same-sign dileptons to increase their signal sensitivity. The distribution of $N_b$ with zero and non-zero lepton(s) in final states are shown in Fig.~\ref{fig:NUMB}:

\begin{figure}[t]
\captionsetup{justification=raggedright,
singlelinecheck=false}
\begin{center}
\begin{subfigure}[b]{0.4\textwidth}
\includegraphics[width=\textwidth]{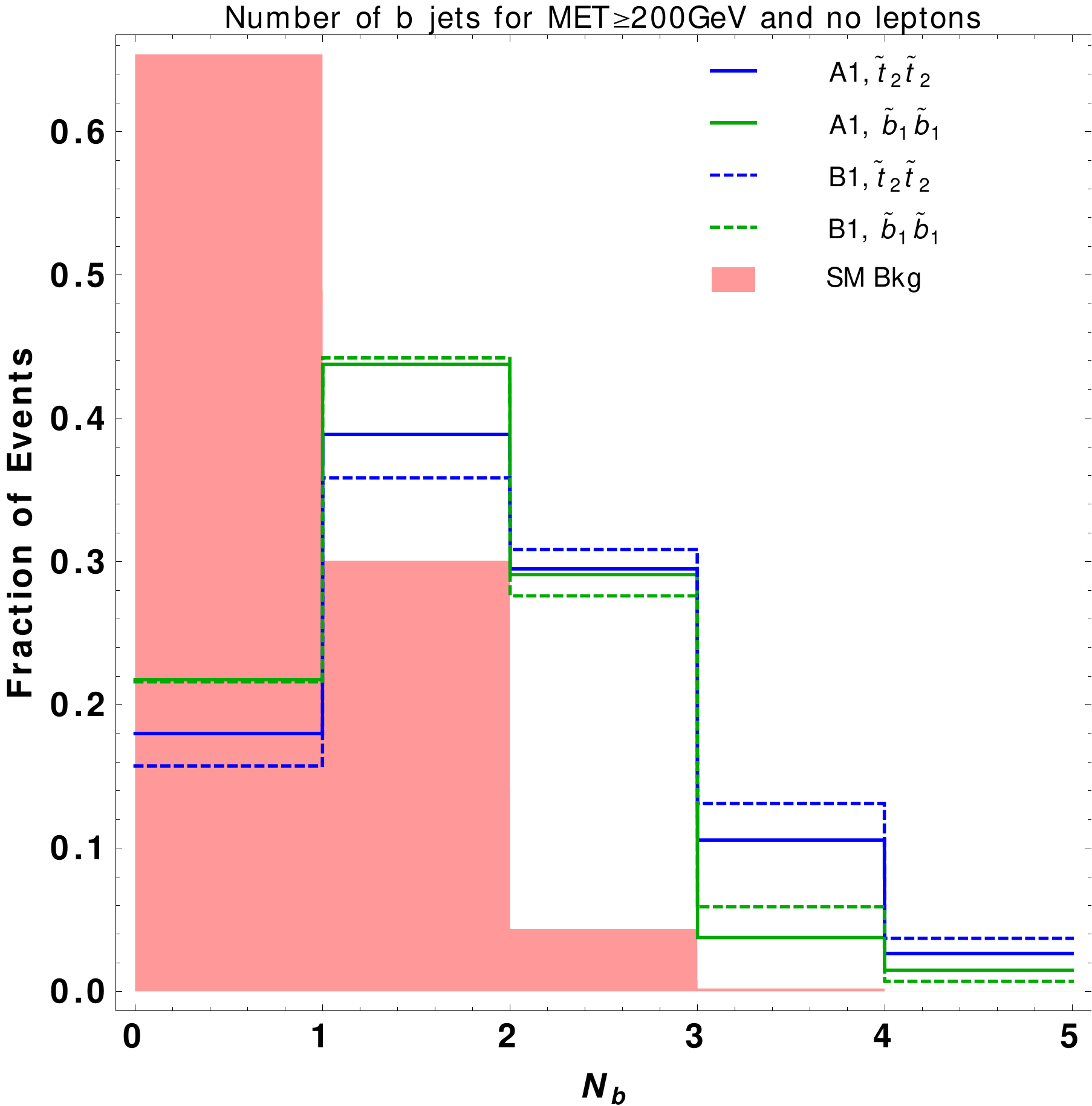}
\end{subfigure}
\begin{subfigure}[b]{0.4\textwidth}
\includegraphics[width=\textwidth]{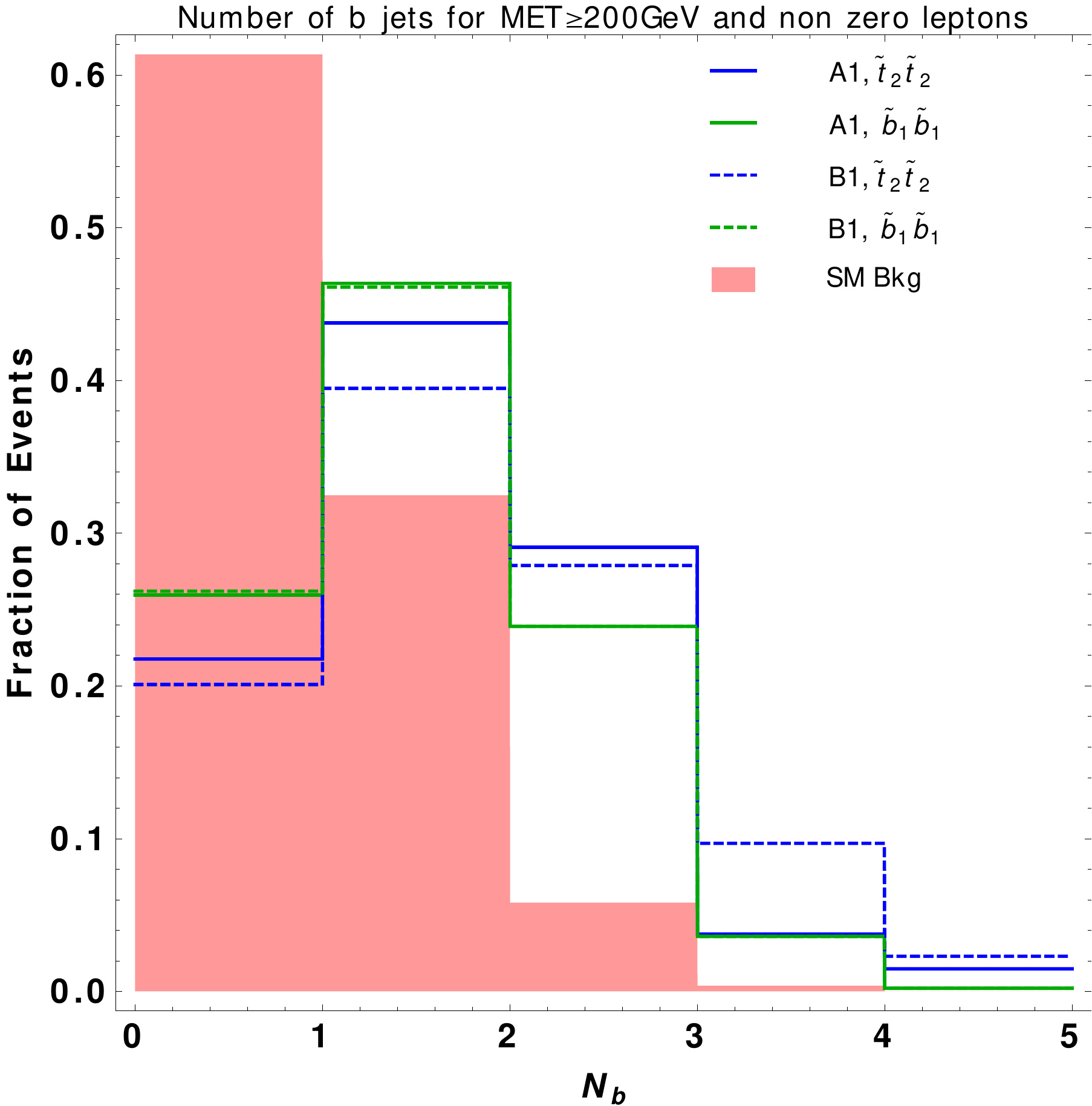}
\end{subfigure}
\caption{Event distribution binned by $N_b$ for both signals and SM background with $\slashed{E}_T \geq 200$~GeV. \textbf{Left:} $N_b$ distribution for events without leptons. The red shade is the SM background. The blue (green) lines are $\ttwo$ ($\bone$) contributions, and the solid (dashed) lines are for the A1 (B1) benchmark model.
\textbf{Right:} the same distributions for events with at least one lepton.}
\label{fig:NUMB}
\end{center}
\end{figure}

Based on these considerations we have tried various search channels by $N_\ell$ and $N_b$. The set of useful signal channels that we found are classified as:

\begin{enumerate}

\item No lepton and no less than 3 $b$ jets with a large MET and no less than 6 jets in total ($0\ell3b$). 

\item 1 lepton and no less than 3 $b$ jets  and no less than 6 jets in total ($1\ell3b$). 

\item 2 opposite-sign (OS) leptons forming a $Z$ candidate with no less than 2 $b$ jets, and 5 or more jets in total ($Z2b$).

\item 2 same-sign (SS) leptons with at least one $b$ jet, also no less than 5 jets in total ($SS+nb$).

\item 3 or more leptons and at least one $b$ jets, the number of jets in total $\geqslant$ 2 (Multi-$\ell$).
\end{enumerate}

Different channels are classified as such that for zero or one lepton channels, more $b$ jets and total number of jets are required to optimize the sensitivity. As the lepton multiplicity increases, we loosen up the requirements on $N_b$, $N_j$ and MET in order to keep more signal events. Some of the channels are essentially the same as those which are already used in experimental $\ttwo$ searches. Others are also close to some other SUSY or new physics searches but have not been applied to $\ttwo$ searches. We also identified some additional requirements which may enhance the signal and background discrimination in some channels. For each signal channel, it would be beneficial to further divide into signal regions based on various energy distributions to utilize the possible different distributions between signals and backgrounds. This requires more sophisticated event simulations to produce accurate event distributions. For simplicity, here we will treat each signal channel as a whole and leave more detailed analyses for the experimental collaborations.

\subsubsection{No lepton, large MET, with three or more $b$ jets ($0\ell3b$)}
In this channel, we require no isolated lepton in the final states,  $N_b \geqslant 3$, and the total number of jets $\geqslant6$. For such fully-hadronic events to be triggered, we require that the leading jet has $p_T \geqslant 250$~GeV and two subleading jets have $p_T \geqslant 90$~GeV. For $b$ jets we also require at least two $b$ jets to have $p_T \geqslant 90$~GeV and the rest of $b$ jets to have $p_T>30$~GeV.

\begin{figure}[t]
\captionsetup{justification=raggedright,
singlelinecheck=false}
\begin{center}
\includegraphics[scale=0.6]{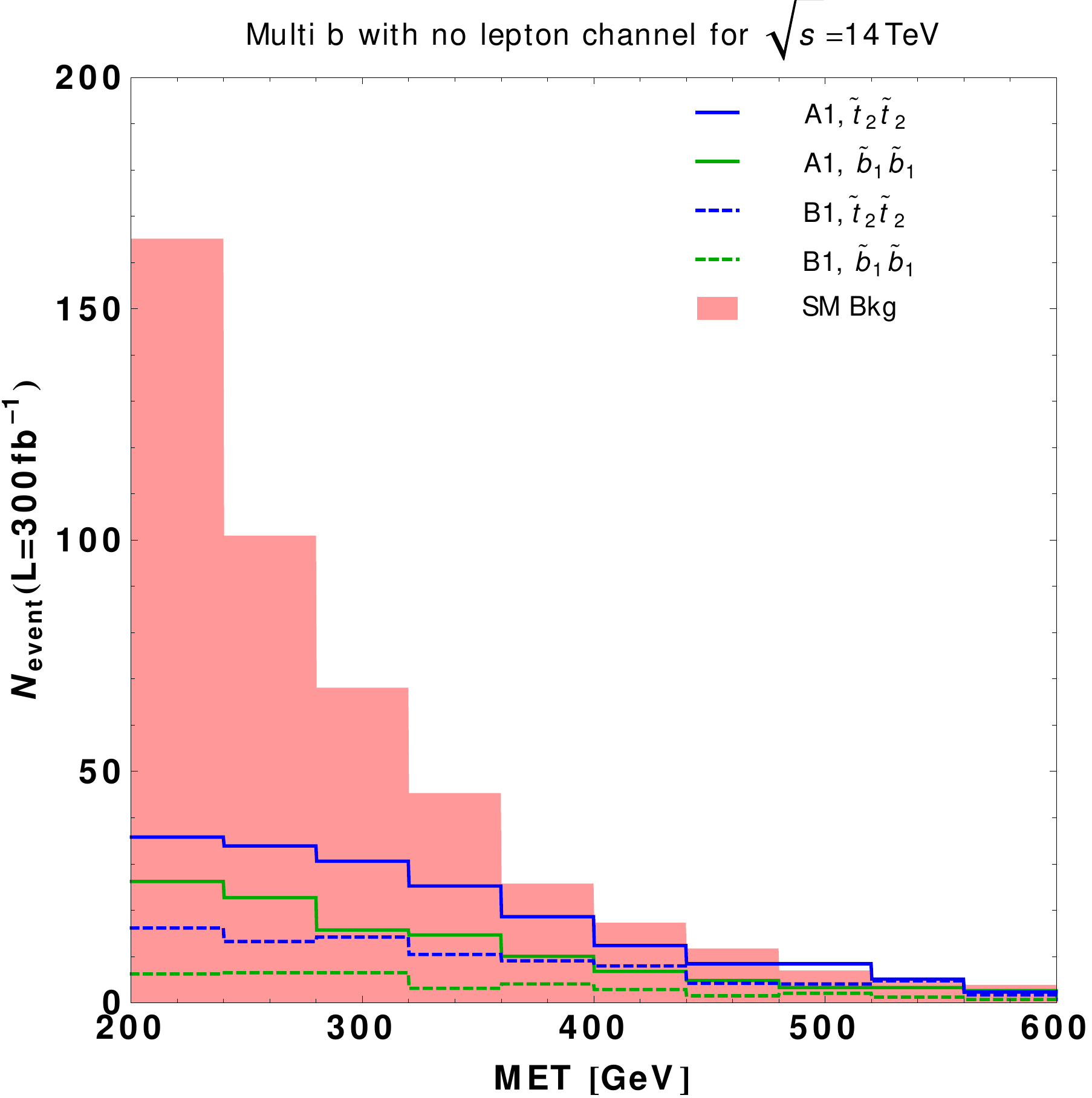}
\caption{The $\slashed{E}_T$ distributions of the $0\ell3b$ channel for benchmarks A1, B1 and the SM  background. 
The red shade is the SM background. The blue lines are $\ttwo$ contributions and the green lines are $\bone$ contributions. The solid lines are for A1 and the dashed lines are for B1.\label{fig:0l3b}
}
\end{center}
\end{figure}

A large $\slashed{E}_T$ cut is imposed in this channel to greatly suppress the contribution from QCD background, allowing us to utilize the Snowmass 2013 Energy Frontier Simulation results. The $\slashed{E}_T$ distributions for some benchmark signals and the background are shown in  Fig.~\ref{fig:0l3b}. We can see that both $\bone\bone$ and $\ttwo\ttwo$ events on average have higher $\slashed{E}_T$ than background events. We impose $\slashed{E}_T >280$~GeV for event selections in this channel.

\begin{table}[ht]
\captionsetup{justification=raggedright,
singlelinecheck=false}
\begin{center}
\captionsetup{justification=raggedright,
singlelinecheck=false}
\begin{tabular}{|c|c|c|c|c|c|c|c|c|c|c|c|c|c|c|c|c|c|c|c|}
\hline\hline
Bkg Total &Model & A1& A2&A3&A4&A5&A6&B1&B2&B3&B4\\\hline
\multirow{2}{*}{191.4}&$\ttwo\ttwo$&119.9&51.2&24.9&7.6&11.8 &33.2 &61.5 &57.6 &26.1 &51.6  \\
&$\bone\bone$&67.0&56.4&20.5&5.0&5.2 &7.6  &25.7 &86.2 &24.6 &31.1 \\\hline

\end{tabular}
\end{center}
\caption{Background and signal events for $0\ell3b$ channel normalized to 300 fb$^{-1}$ integrated luminosity at 14 TeV LHC for each benchmark model point.\label{table:0l3b}}
\end{table}

The number of signal events of each benchmark model and the background events passing the selection in this channel for a 300 fb$^{-1}$ integrated luminosity is listed in Table~\ref{table:0l3b}. We can see that the $0\ell3b$ channel can be useful for both $\ttwo$ and $\bone$ searches. Na\"ively one might expect that this channel is more useful for the $\ttwo$ search than the $\bone$ search because the $Z$ or $h$ bosons from $\ttwo$ decays give the extra $b$ jets. This is true for some (e.g., $\ttwo$ has a large branching ratio in decaying to $h/Z$ in A1, A5, A6, which produces more $b$ jets on average) but not all of the benchmark models. For example, in some Type B models (in particular B2), $Z$ or $h$ bosons can also arise from the cascade decays of $\bone$ through heavier neutralinos, so $\bone$ can provide comparable number of events as  $\ttwo$.  Even though in Type A models $\bone\bone$ pair production can only give two $b$ jets at the parton level, additional $b$-tagged jets can arise from QCD radiations and mis-tagged light jets. In addition, $\bone$ has a larger production cross section than $\ttwo$ because it is lighter. As a result, the number of events from $\bone$ and from $\ttwo$ can be comparable in many Type A models, and this channel can also be sensitive to $\bone$ pair production. Among the heavier benchmarks A4-A6, this channel is most sensitive to A6 and its contribution mainly comes from $\ttwo$ because of the large $\ttwo \to h \tone$ branching ratio.

For Type B models, B3 has fewer signal events due to the more compressed spectrum which results in softer final state particles and hence a lower signal efficiency. This is also true for other signal channels discussed later. For B4, even though the production cross sections of $\ttwo$ and $\bone$ pairs are small, the final state particles are harder due to the large mass splittings. The signal efficiency is better. In addition, there are significant branching ratios of $\ttwo$ and $\bone$ decaying to $tbb$ and $ttb$ which also give a higher number of $b$ jets.

\subsubsection{One lepton with three or more $b$ jets ($1\ell3b$)}
\begin{figure}[t]
\begin{center}
\includegraphics[scale=0.6]{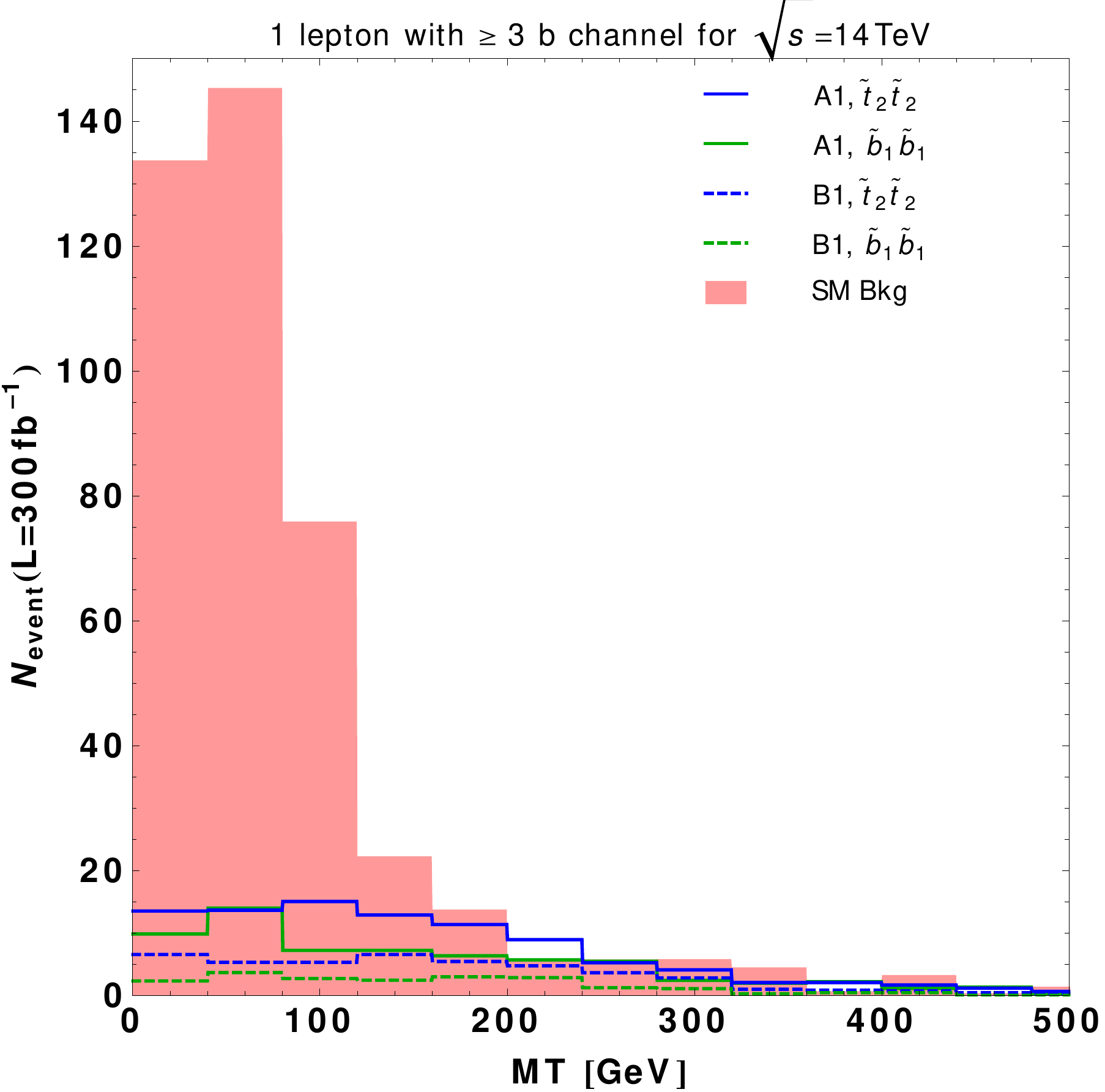}
\caption{$M_T$ distributions for of $1\ell3b$ channel.}
\label{fig:1l3b}
\end{center}
\end{figure}

In this channel, the isolated lepton is required to have $p_{T\ell}\geqslant25$~GeV. We also require more than 5 jets with $p_{T}\geqslant25$~GeV for each jet. The leading $b$ jet should have $p_{Tb}\geqslant 90$~GeV and other $b$ jets have  $p_T>30$~GeV. The $\slashed{E}_T$ is required to be $\geqslant 200$~GeV. For the SM backgrounds, a significant $\slashed{E}_T$ can arise due to the neutrino produced by $W\to \ell\nu$ decay. A cut on the transverse mass $M_T=\sqrt{2(p_{T\ell}\slashed{E}_T)(1-\cos\Delta\phi)}$ can be an effective way to eliminate most of the SM backgrounds, since $M_T$ from single $W$ leptonic decay would have a drop off around $m_W$. The $M_T$ distributions for signals and the background for the single lepton channel are shown in Fig.~\ref{fig:1l3b}. We impose a $M_T>160$~GeV cut to enhance the signal significance.

\begin{table}[ht]
\captionsetup{justification=raggedright,
singlelinecheck=false}
\begin{center}

\begin{tabular}{|c|c|c|c|c|c|c|c|c|c|c|c|c|c|c|c|c|c|c|c|}
\hline\hline
Bkg Total &Model & A1& A2&A3&A4&A5&A6&B1&B2&B3&B4\\\hline
\multirow{2}{*}{42.5}&$\ttwo\ttwo$&39.9&23.8 &10.8 &3.8 &3.2 &6.2  &20.2 &14.6 &7.8 &13.1 \\
&$\bone\bone$&27.1&13.5 &8.9 &2.6 &2.4 &3.1  &9.3 &1.3 &3.0 &12.5 \\\hline
\end{tabular}
\end{center}
\caption{Background and signal events for the $1\ell3b$ channel with a 300 fb$^{-1}$ integrated luminosity for each model point.\label{table:1l3b}}
\end{table}

This is one of the main channels for current experimental $\ttwo$ searches~~\cite{Aad:2015pfx,Khachatryan:2014doa}. From Table~\ref{table:1l3b} we see that $\bone$ could also give nearly comparable contributions for many benchmark points (except B2 where $W$'s are not produced in $\bone$ decays), due to the larger cross section for being lighter.

\subsubsection{Opposite-sign dilepton (Z) channels}
Events with an OS lepton pair and $b$ jets would be dominated by SM $t\bar{t}$ production. Here we define the OS lepton channels by having two OS leptons with $p_T>15$~GeV, with $\slashed{E}_T>150$~GeV. 
Also each event is required to have $\geqslant$ 2 $b$-jets with $p_T>30$~GeV and $\geqslant$ 5 jets  of $p_T>25$~GeV in total. A rough estimation for this channel predicts $\sim 800$ background events and $\lesssim 10$ signal events for those heavy benchmark points such as A4-A6 or B2-B4, yielding a significance too low to be useful. Therefore, additional requirement is needed to suppress the background and we focus on the case where the lepton pair come from the $Z$ decay, since there is always a significant branching ratio of producing $Z$'s in $\ttwo$ decays. In addition, $Z$ can also arise from heavier neutralino decays in Type B models. To be identified as a $Z$, the OS leptons of the same flavor is required to have $\Delta R <1.5$ and an invariant mass $|m_{\ell\ell}-m_Z|<10$~GeV. 

The $Z2b$ is also a standard search channel for the $\ttwo$ pair production~\cite{Aad:2015pfx,Aad:2014mha,Khachatryan:2014doa}. There is also an attempt to explain the ATLAS $Z$+jets+MET excess~\cite{Aad:2015wqa} by the two mixed stop system~\cite{Collins:2015boa}. Here we notice that most signal events contain additional $W$ or $Z$ bosons. Some of them may be boosted if they come from the decay of a heavy particle. We therefore explore the possibility that additional vector tags may be helpful with the signal and background discrimination. We divide the events passing the above requirements into exclusive channels with no vector tag ($Z2b$) and with at least one vector tag ($Z2bV$). They are listed in Table~\ref{table:Z2bV}. We see that the channel with a vector tag in general does better than without the vector tag,
especially for model points with larger $\mttwo-\mtone$ and higher BR$(\ttwo \to Z\tone)$ (e.g., A1, A6 and B4). Since the vector-tagging technique we used here is rather crude, further improvements with advanced vector-tagging techniques are possible.

These channels are in general more useful for the $\ttwo$ search, because in most benchmark models, $Z$ is rarely produced in $\bone$ decays. The exceptions are B2 and B3, where $Z$ can be produced from the neutralino decays. However, the signal efficiency for B3 is small due to its compressed spectrum. Only for B2 $\bone$ can give a comparable contribution to $\ttwo$.

\begin{table}[ht]
\captionsetup{justification=raggedright,
singlelinecheck=false}
\begin{center}
\begin{tabular}{|c|c|c|c|c|c|c|c|c|c|c|c|c|c|c|c|c|c|c|c|}
\hline\hline
Channel&Bkg Total &Model & A1& A2&A3&A4&A5&A6&B1&B2&B3&B4\\\hline
\multirow{2}{*}{$Z2bV$}&\multirow{2}{*}{16.7}&$\ttwo\ttwo$&27.1&6.5&3.0&1.2&1.5&2.8&9.2&3.2&2.0&2.4 \\
& &$\bone\bone$&1.5&0.0&0.2&0.0&0.0&0.0&0.5&3.1&0.3&0.1 \\\hline
\multirow{2}{*}{$Z2b$}&\multirow{2}{*}{29.2}&$\ttwo\ttwo$&20.5&7.0&2.7&0.9&1.4&1.2&9.0&2.7&1.6&1.0 \\
& &$\bone\bone$&1.7&0.0&0.5&0.1&0.0&0.0&0.5&3.5&0.3&0.1 \\\hline
\end{tabular}
\end{center}
\caption{Background and signal events for $Z2bV$ and $Z2b$ channel for each model point with a 300 fb$^{-1}$ integrated luminosity.}
\label{table:Z2bV}
\end{table}

The significance of this channel may be further improved if one can suppress the fake $Z$ bosons made of two leptons coming from opposite sign $W$'s from the SM $t\bar{t}$ background. This contamination may be estimated by the opposite-sign, opposite-flavor dilepton events which satisfy the invariant mass and $\Delta R$ requirements, after taking into account the different efficiencies of the electron and the muon. Here instead we introduce a simple kinematic variable dubbed ``leverage.'' It is inspired by $M_T$ and can be considered as a generalization applying to more than one leptons (or even other final state particles) together with MET.  With multiple leptons it is defined as
\begin{equation}
L_\ell=\left(p_T^{\rm miss}
{\sum} (1-\cos \Delta \phi_i)\right)/N_\ell
\end{equation}

\begin{figure}[t]
\captionsetup{justification=raggedright,
singlelinecheck=false}
\begin{center}
\includegraphics[scale=0.35]{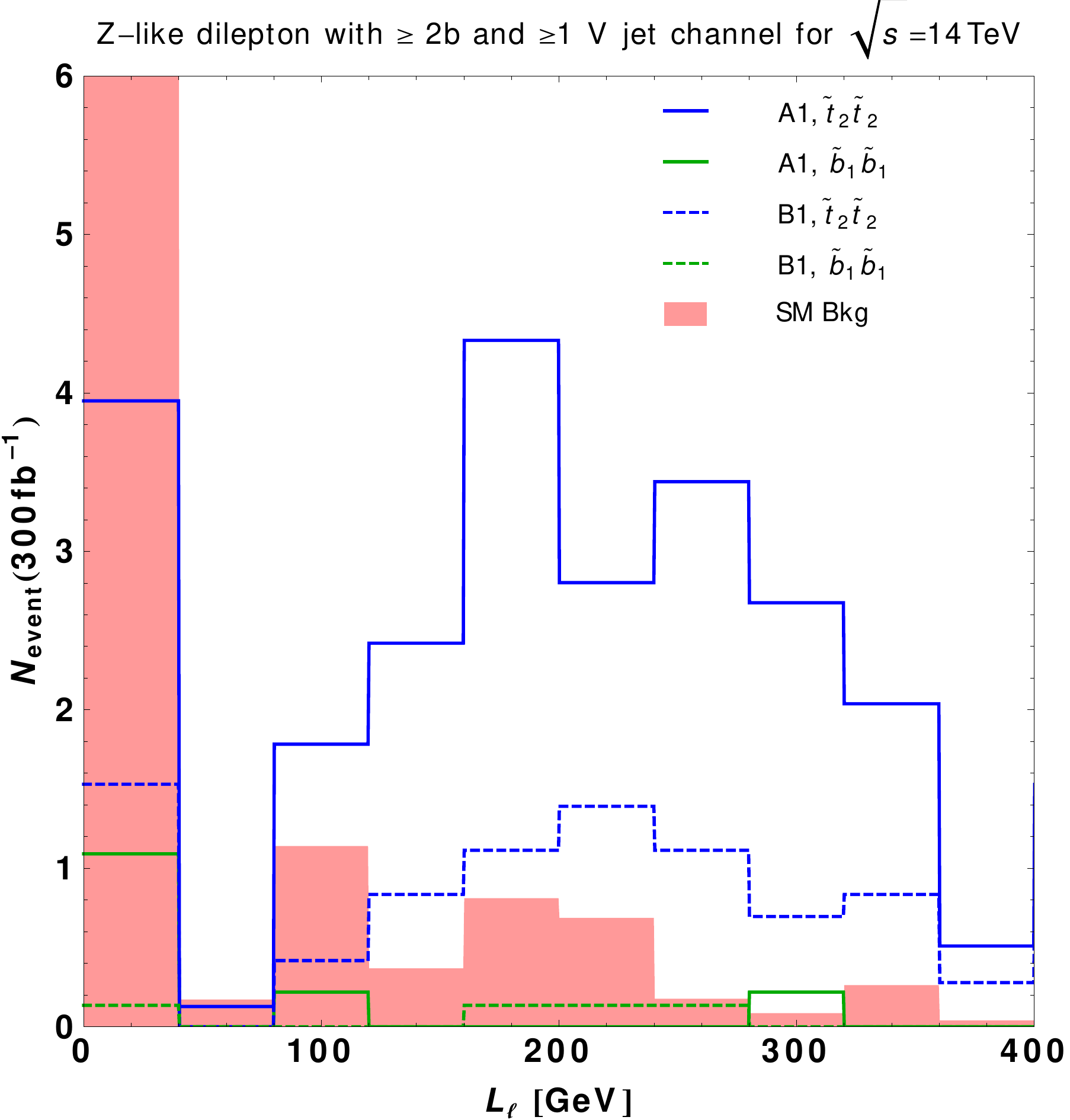}
\includegraphics[scale=0.35]{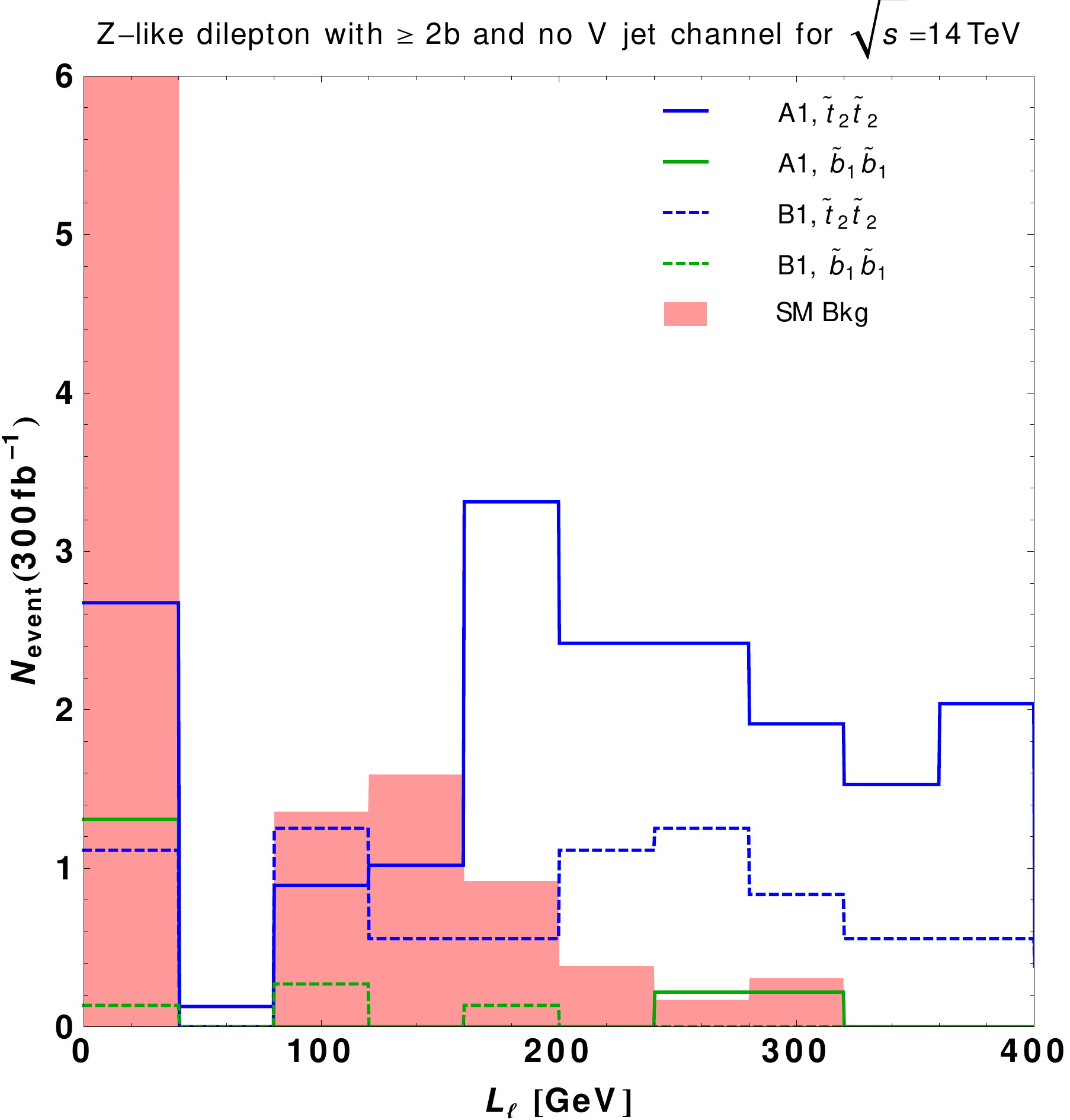}
\caption{Histogram of leptonic $Z$ with  $\geqslant 2\, b$-jets and $
\geqslant 5$ jets binned by $ L_{\ell}$. \textbf{Left:} With at least one extra vector-tagged jet. \textbf{Right:} without vector-tagged jets.}
\label{fig:Zbb}
\end{center}
\end{figure}

In the SM fake $Z$ events, since we require $\Delta R\leqslant1.5$, the MET given by two neutrinos from $W$ decay would tend to be in the same direction as the fake $Z$ direction. On the contrary, there is much less such correlation for SUSY events, since $Z$ produced by $\ttwo$ decay can have a different direction from the MET. As it can be seen in both plots in Fig.~\ref{fig:Zbb}, most of the background events have a small $L_{\ell}$ and a suitable cut on $L_{\ell}$ can increase both the significance and $S/B$ ratio effectively. To enhance the signal significance, we put an additional $L_{\ell}>40$~GeV cut for both $Z2bV$ and $Z2b$ channels besides the cuts aforementioned. The numbers of events after the cut are listed in Table~\ref{table:L_cut}. We can see that it significantly reduces the remaining background events while retaining most of the signal events.

\begin{table}[ht]
\captionsetup{justification=raggedright,
singlelinecheck=false}
\begin{center}
\begin{tabular}{|c|c|c|c|c|c|c|c|c|c|c|c|c|c|c|c|c|c|c|}
\hline\hline
Channel&Bkg Total &Model & A1& A2&A3&A4&A5&A6&B1&B2&B3&B4\\\hline
\multirow{2}{*}{$Z2bV$}&\multirow{2}{*}{4.0}&$\ttwo\ttwo$&23.1&5.2 &2.7 &0.9 &1.3 &2.3  &7.6 &2.7 &1.9 &2.1 \\
& &$\bone\bone$&0.4 &0.0&0.2 &0.0 &0.0 &0.0 &0.4 &1.3 &0 &0.1  \\\hline
\multirow{2}{*}{$Z2b$}&\multirow{2}{*}{4.9}&$\ttwo\ttwo$&17.8 &5.9&1.9 &0.9 &1.2 &1.0 &7.9 &2.2 &1.1 &0.9  \\
& &$\bone\bone$&0.4 &0.0&0.0 &0.1 &0.03 &0.03  &0.4 &3.1 &0.2 &0.0  \\\hline
\end{tabular}
\end{center}
\caption{Background and signal events for $Z2bV$ and $Z2b$ channel for each model point after the $L_{\ell}>40$~GeV cut with a 300 fb$^{-1}$ integrated luminosity.}
\label{table:L_cut}
\end{table}

\subsubsection{Same-sign dilepton channel}
The previous channels are sensitive to the $\ttwo \to \tone+ Z/h$ decays which are the focus of existing experimental analyses based on the simplified model approach. However, as we see in the benchmark models, it is common to have large fractions of final states with high multiplicities of $W$ bosons from both $\ttwo$ and $\bone$ decays. It is therefore important to study signals from multiple $W$'s in $\ttwo$ and $\bone$ searches. A very useful signal for multiple $W$'s is the same-sign dilepton which is relatively rare in the SM.
For this study we define the same-sign dilepton channel to have two leptons of the same charge and $p_T>15$~GeV, with $\slashed{E}_T >190$~GeV for each event. Also, we require at least one $b$-jet with $p_T>30$~GeV and the total number of jets of $p_T>25$~GeV is required to be more than 4. We further divide these events into two channels: $SS1b$ for events with exactly one $b$-tagged jet and $SS2b$ for events with two or more $b$-tagged jets.

\begin{figure}[t]
\captionsetup{justification=raggedright,
singlelinecheck=false}
\begin{center}
\includegraphics[scale=0.4]{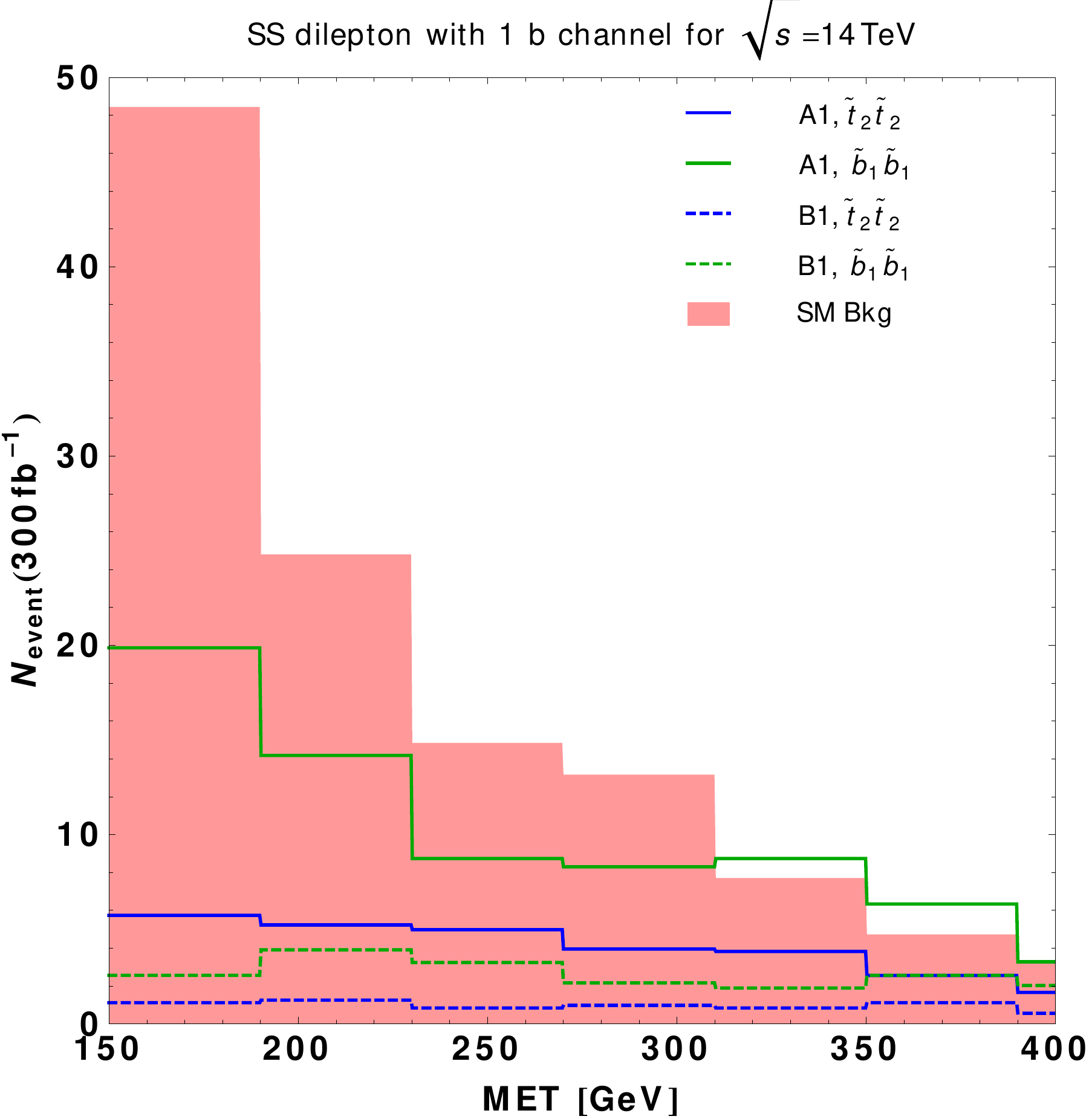}
\includegraphics[scale=0.4]{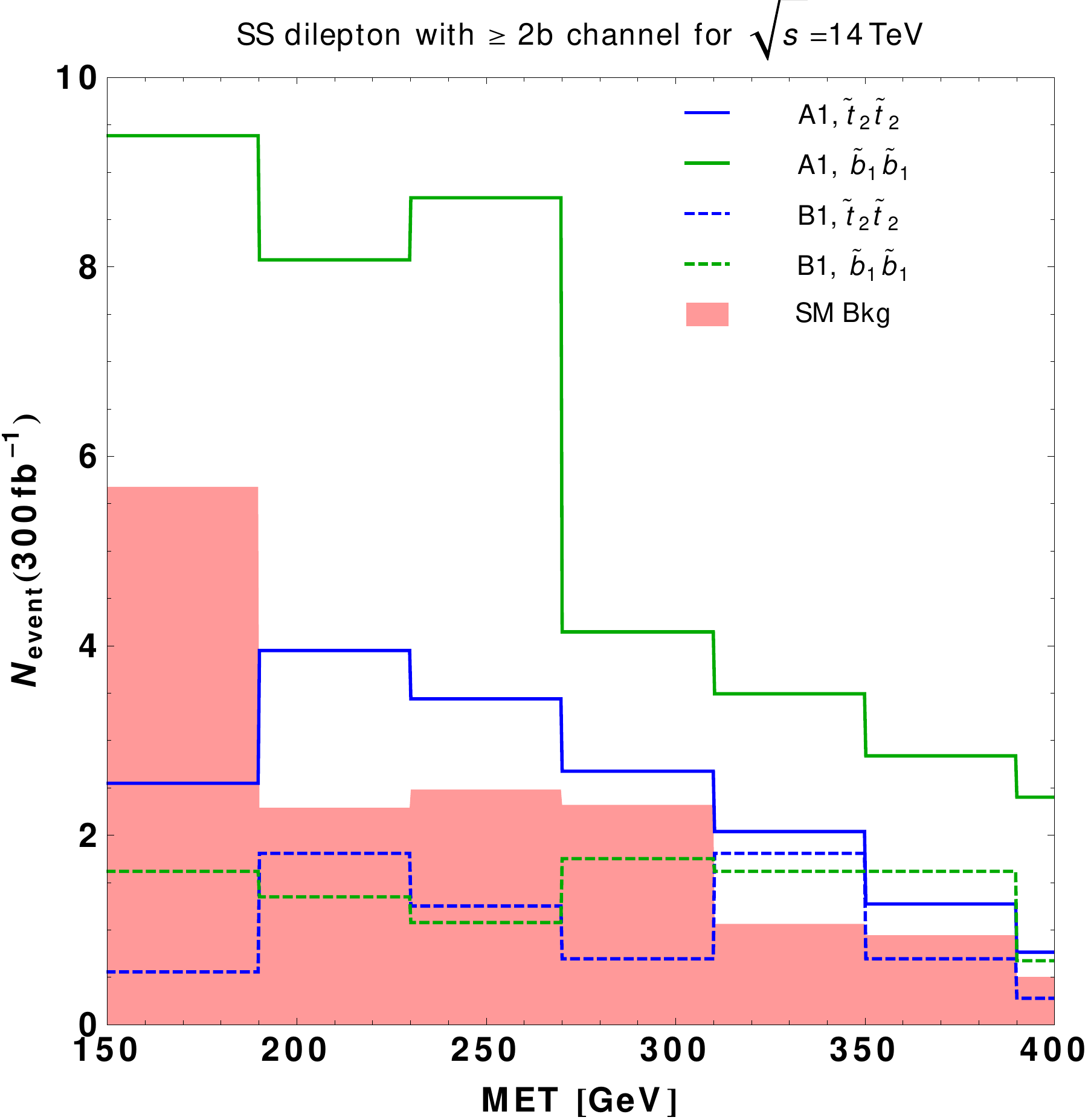}

\caption{The $\slashed{E}_T$ distributions of the same-sign lepton pair with \textbf{Left:} 1 $b$ jet and \textbf{Right:} more than 1 $b$ jets.}
\label{fig:SS}
\end{center}
\end{figure}

Ignoring the contribution of misidentified leptons, only a few SM processes can generate a pair of same-sign dilepton, such as multiple vector boson or $t\bar{t}+$ boson production, etc. In our case, the dominant background is $t\bar{t}W/t\bar{t}Z/t\bar{t}h$ production. However, these processes' contribution suffer from small production cross sections ($\sim 3$~pb in total). The distributions of the signal and background events for the $SS1b$ and $SS2b$ channels in $\slashed{E}_T$ are shown in Fig.~\ref{fig:SS}. We see that our $\slashed{E}_T$ cut can further reduce the SM backgrounds.

\begin{table}[t]
\captionsetup{justification=raggedright,
singlelinecheck=false}
\begin{center}
\begin{tabular}{|c|c|c|c|c|c|c|c|c|c|c|c|c|c|c|c|c|c|}
\hline\hline
Channel&Bkg Total &Model & A1& A2&A3&A4&A5&A6&B1&B2&B3&B4\\\hline
\multirow{2}{*}{$SS1b$}&\multirow{2}{*}{76.6}&$\ttwo\ttwo$&26.2&20.9 &10.1 &3.0 &1.1 &1.2  &7.1 &2.0 &2.6 &1.4  \\
& &$\bone\bone$&62.0&29.8 &16.2 &3.4 &3.9 &5.8 &19.5 &0.0 &1.9 &4.5  \\\hline
\multirow{2}{*}{$SS2b$}&\multirow{2}{*}{10.8}&$\ttwo\ttwo$&16.8&15.5 &7.1 &2.1 &1.0 &1.1 &7.4 &2.0 &3.4 &2.9  \\
& &$\bone\bone$&34.9&15.1 &8.2 &2.1 &1.7 &3.5  &11.5 &0.0 &1.0 &5.3 \\\hline
\end{tabular}
\end{center}
\caption{Background and signal events for $SS1b$ and $SS2b$ channel for each model point with a 300 fb$^{-1}$ integrated luminosity.\label{table:ss2l}}
\end{table}

One can see from Table~\ref{table:ss2l} that the numbers of signal events are comparable for $SS1b$ and $SS2b$ channels. However, $SS1b$ has much more background events due to mis-tagged $b$ jets, so $SS2b$ is expected to have better reaches. In general, for benchmark points where $\bone$ has a large branching ratio of decaying to $\tone W$ (i.e., other than B2 and B3), more signal events come from $\bone$ due to the larger production cross section. Nevertheless, $\ttwo$ can also give a significant contribution. As $\ttwo$ pair decays may produce more than 4 $W$'s in the final states, the probability of getting same-sign dileptons could be helped by the combinatorial factor.  Also, in models where the mass difference between $\bone$ and $\tone$ is small, the leptons coming from $\bone$ are softer and hence have a lower signal efficiency. For example, A1 and A2 have similar production rate for $\bone$ pairs, but the signal efficiency of the latter is almost halved because of its small $\bone$--$\tone$ mass difference. In this case, the contributions from $\ttwo$ and $\bone$ can even be comparable.

The same-sign dilepton signals, although not used in $\ttwo$ search yet, have been applied to many other new physics searches. Interestingly, excesses in the same-sign dilepton channel with $b$-jets and MET are found in both Run 1 and Run 2 of LHC in many separate analyses by the ATLAS and CMS collaborations. These include CMS SUSY Search~\cite{Chatrchyan:2013fea}, ATLAS SUSY Search~\cite{Aad:2014pda}, CMS $tth$ Search~\cite{CMS:2013tfa}, ATLAS Exotica Search~\cite{Aad:2015gdg}, and ATLAS $tth$ Search~\cite{Aad:2015iha} in Run 1 and  ATLAS $tth$ Search~\cite{ATLAS:2016ldo}, CMS $tth$ Search~\cite{CMS:2016vqb} in Run 2. On the other hand, there are no significant excesses in Run 2 SUSY searches~\cite{ATLAS:2016kjm,CMS:2016vfu}. The SUSY analyses were based on the simplified model of sbottom decay $\bone \to t +(\cone^\pm \to W^\pm +\cone^0)$. However, the branching ratio of such a decay chain is never close to 100\% due to the presence of other decays ($\bone \to b +\tilde{\chi}_2^0 \mbox{ or } \cone^0$), and these other decays generally give stronger constraints as we saw in the previous section. Ref.~\cite{Huang:2015fba} proposed to explain the excesses from the right-handed stop production with the decay $\tilde{t}_R \to t+ (\tilde{B} \to W^\pm + \tilde{W}^\mp)$ where $\tilde{W}^\mp$ is closely degenerate with the neutral Wino which is assumed to be the LSP, and hence its decay products are too soft to be seen. With a suitable arrangement of the spectrum, the branching ratio of this decay chain can be close to 100\%, and the excesses could be explained by a $\tilde{t}_R$ of $\sim 550$~GeV.

The spectrum in our study provides an alternative way to explain the excesses. From Table~\ref{table:A_final}, we can see that the sbottom pair decays to the $2t+2W+\slashed{E}_T$ final state which gives the desired signals are close to 100\% in Type A models. In addition, a substantial fraction of $\ttwo$ pairs also give  $ttWW+\slashed{E}_T+X$ final states which contribute to the signal. We expect that our benchmark models with light spectra may produce same-sign dilepton events compatible with the excesses observed in experiments. To minimize the systematic uncertainties in comparing with experimental excesses, we follow Ref.~\cite{Huang:2015fba} to normalize the signal strength of our benchmark points to the SM $t\bar{t}h$ signal strength,  then compare the simulation results at 13 TeV to the best fit signal strengths of the new Run 2 results: $\mu=4.0^{+2.1}_{-1.7}$ of the ATLAS $2l0\tau_{had}$ signal region~\cite{ATLAS:2016ldo} and $\mu=2.7^{+1.1}_{-1.0}$ of the CMS 2LSS category for $t\bar{t}h$ searches~\cite{CMS:2016vqb}. For our benchmark model A1 (A2), we get a total signal strength of $\mu=2.4 (1.8)$, with $\mu_{\bone}=1.1 (0.6)$ from $\bone$ pairs and $\mu_{\ttwo}=0.3 (0.2)$ from $\ttwo$ pairs. They are in the ballpark of the observed excesses in Run 2. Furthermore, we also check our benchmark points with the 95\% CL upper limits on the number of SS2L events in the ATLAS SUSY search~\cite{ATLAS:2016kjm}. For our A1 (A2), we get 8.0(4.8)/0.9(0.4) events in the corresponding SR1b/SR3b signal regions where the observed 95\% CL upper limits are 10.3/4.9 .

\subsubsection{Channels with multiple leptons}
Multiple $W$'s and $Z$'s also give rise to multilepton signals.
For multilepton channels, we require 3 or more leptons with $p_T>15$~GeV and $\slashed{E}_T>100$~GeV in the final state. Since the chance of finding more than two leptons in an event is rare, we loosen the $N_b$ cut to $\geqslant 1$ and $N_j\geqslant 2$ in these cases. Each $b$ jet is required to have $p_T>30$~GeV while for other jets $p_T$ should be greater than 25~GeV. The multilepton events are then split into two different channels based on the number of $Z$-like lepton pairs. We ascribe those events with at least one $Z$-like lepton pair to the $\ell Zb$ channel, and the rest are recorded by the $3\ell b$ channel.

\begin{figure}[t]
\captionsetup{justification=raggedright,
singlelinecheck=false}
\begin{center}
\includegraphics[scale=0.35]{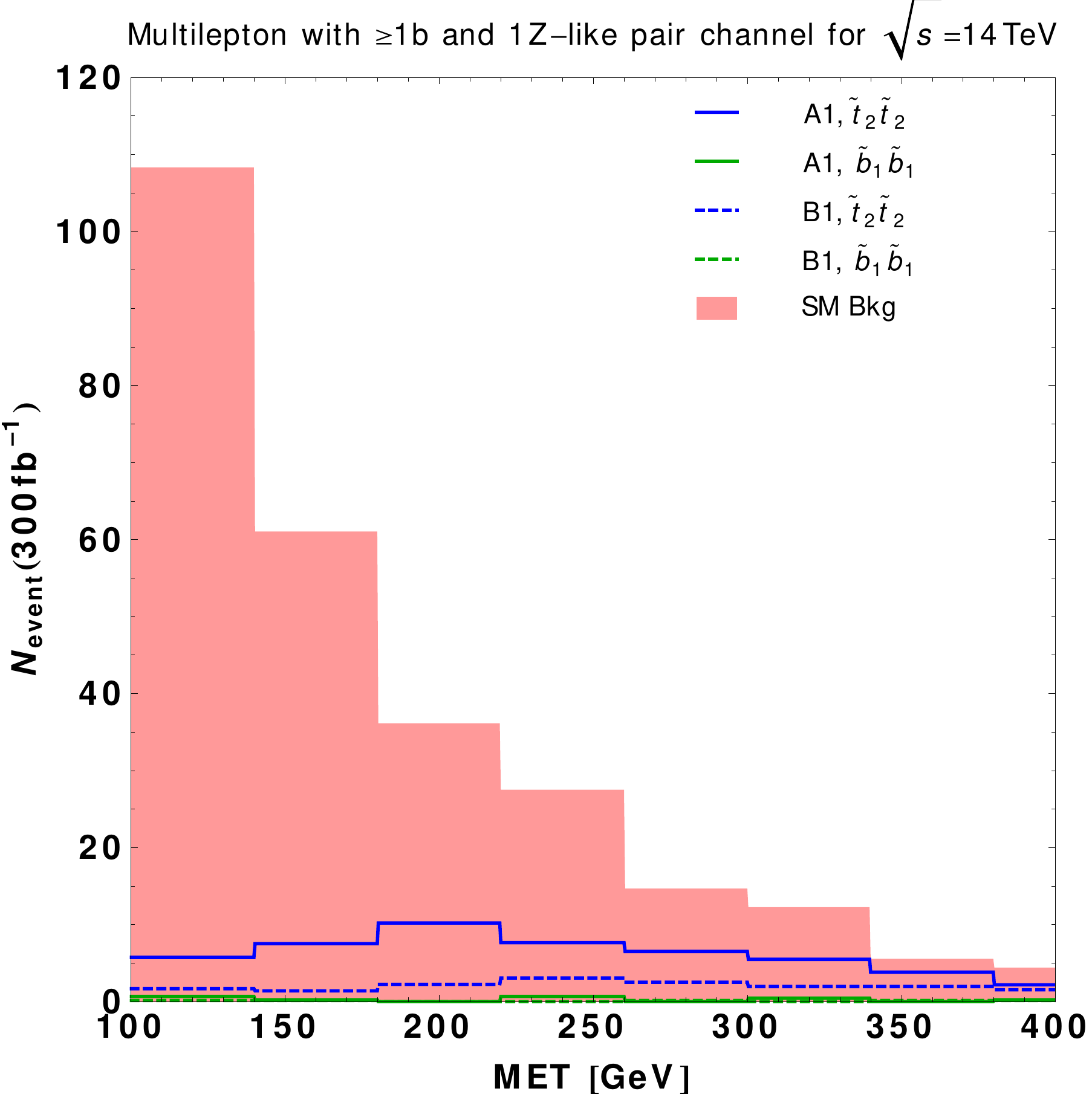}
\includegraphics[scale=0.35]{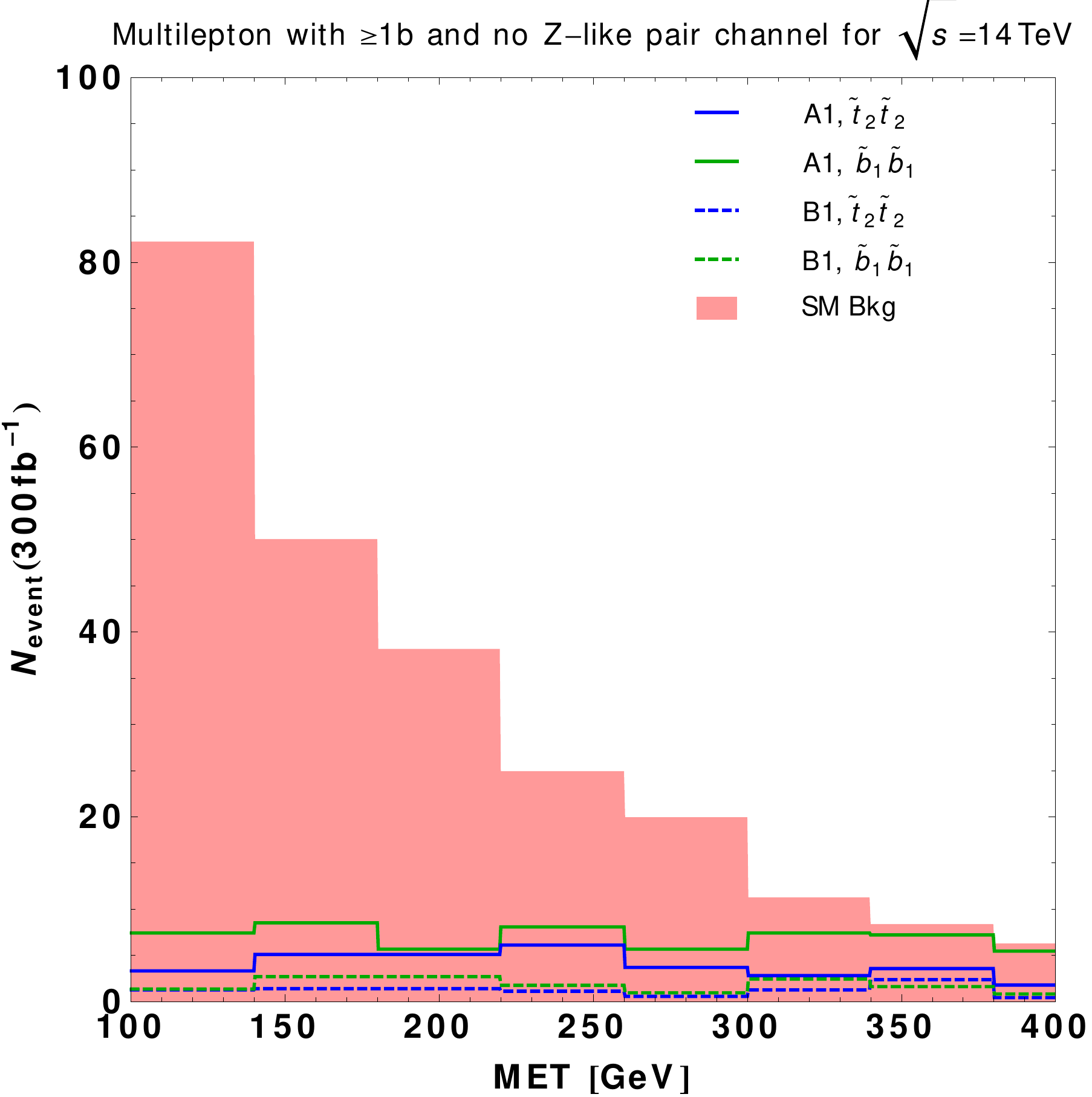}
\caption{Histograms in $\slashed{E}_T$ of multiple lepton channels. We require $\geqslant1\, b$ jet and $\geqslant2$ jets in total. \textbf{Left:} Channel with at least 1 $Z$ like lepton pairs and extra lepton(s). \textbf{Right:} Channel without any $Z$ like lepton pair.}
\label{fig:multilep}
\end{center}
\end{figure}

\begin{table}[ht]
\captionsetup{justification=raggedright,
singlelinecheck=false}
\begin{center}
\begin{tabular}{|c|c|c|c|c|c|c|c|c|c|c|c|c|c|c|c|c|c|}
\hline\hline
Channel&Bkg Total &Model & A1& A2&A3&A4&A5&A6&B1&B2&B3&B4\\\hline
\multirow{2}{*}{$\ell Z b$}&\multirow{2}{*}{279.3}&$\ttwo\ttwo$&57.1&20.1&9.2&3.0&3.4&4.5&18.2&6.7&5.0&3.2\\
& &$\bone\bone$&3.1&1.6&0.5&0.2&0.2&0.2&0.9&0.9&0.5&0.2\\\hline
\multirow{2}{*}{$3 \ell b$}&\multirow{2}{*}{250.8}&$\ttwo\ttwo$&39.1&27.2&12.8&3.5&1.7&1.7&12.2&2.7&4.0&2.6\\
& &$\bone\bone$&66.6&23.9&13.0&3.2&2.9&4.8&18.6&0.9&1.9&5.0\\\hline
\end{tabular}
\end{center}
\caption{Background and signal numbers for each model point in $\ell Zb$ and $3\ell b$ channels with a 300 fb$^{-1}$ integrated luminosity.}
\label{table:multilep}
\end{table}

The SM backgrounds for multileptons with $b$ jets coming from $t\bar{t}+W/Z/h$ and vector boson pair produced with extra jets. The $\slashed{E}_T$ distributions of signals and backgrounds for both channels are shown in Fig.~\ref{fig:multilep}. We see that there are still substantial backgrounds for both channels. The numbers of background and signal events for the benchmark points with a 300 fb$^{-1}$ integrated luminosity are listed in Table~\ref{table:multilep}. We note that $Z$ bosons are mostly produced in $\ttwo$ decay so $\ell Zb$ signal is $\ttwo$ specific. As a result the $\ell Z b$ channel has less total signal events than the $3\ell b$ channel. Anyway, the large number of background events make either channel less effective for our benchmark models than the previous channels. 

\subsubsection{Other potential channels}

As we discussed, the decay chain $\ttwo \to W\bone \to WW \tone \to WWt \cone^0$ is common for a typical spectrum and has a significant branching ratio in several benchmark models (e.g., A2--A4). In this case there can be up to 6 $W$'s in the final state, which could give rise to same-sign trilepton events. The SM background would be extremely low. The signal event rate is low so one does not expect it to be the discovery channel. However, it may provide useful information about the spectrum after the discovery.

For type B models, the long decay chains may even produce multiple $t$'s in the final states. This occurs in the benchmark B4, where the $\ttwo$ decay can give 3 $t$'s and the $\bone$ decay can produce $ttb$. With 6 tops it can also produce same-sign trilepton events. In addition, specialized searches for multiple top final states such as Ref.~\cite{multitop} may be sensitive to this type of spectrum, although it is not expected to be the first discovery channel either.

\subsection{Significance}

From the tables of signal and background events, we can calculate the signal significances of each channel for the benchmark models. However, there are other effects which can affect the real significances of each channel. First, we need to consider the effect of systematic errors, especially when $S/B$ ratio is small, the uncertainties in the background normalization can overwhelm the signal. There are many factors which could introduce systematic errors, such as ($b$) jet tagging efficiency, detector resolutions, or PDF uncertainties, etc.. Second, if the distributions of the signal and the background on some kinematic variables are very distinct and accurately known, the significance could be enhanced by dividing each channel into several signal regions according to the values of the variables. Here we just use the likelihood method to evaluate the effects of the systematic  errors from the background normalizations. We assume that the overall number of background events respects the normal distribution with a fractional uncertainty $\sigma_{B}\propto B$. The likelihood is defined to be 
\begin{equation}
Q=\frac{\int \mathcal{L}(S+B,S+B')P(B')dB'}{\int \mathcal{L}(S+B,B')P(B')dB'}
\end{equation}
where $S$ and $B$ are corresponding numbers of signal and background events, $\mathcal{L}(x,\mu)=\frac{\mu^{x}e^{-\mu}}{x!}$, and $P(B)$ is the normalized normal distribution with the mean $B$ and a standard deviation $\sigma_B$. The final significance from this method is simply given by $\sqrt{2\log(Q)}$. For the case with no systematic error, $\sigma_B=0$, this equation simply reduces to the standard formula~\cite{Cowan:2010js}:
\begin{equation}
\sigma = \sqrt{\left[2(S+B)\log\left( \frac{S+B}{B} \right)-S\right]}.
\label{eq:significance}
\end{equation}

The systematic uncertainties in general depend on signal channels. Without knowing the exact numbers we will calculate the significances with the assumption of a 10\% uncertainty in background normalization for each channel and compare them with the significances obtained without systematic errors to see their effects.  
The result for Type A and Type B models by combining signal events from both $\ttwo$ and $\bone$ are listed in Table~\ref{table:A_significance} and Table~\ref{table:B_significance} respectively. (Individual significances from $\ttwo$ or $\bone$ can also be easily obtained from the Tables of event numbers in the previous subsection.) The significances for an integrated luminosity different from 300 fb$^{-1}$ can be obtained by a simple rescaling.
\begin{table}[ht]
\captionsetup{justification=raggedright,
singlelinecheck=false}
\begin{center}
\begin{tabular}{|c|c|c|c|c|c|c|}
\hline 
Significance & A1 & A2 & A3 & A4 & A5 & A6\\ 
\hline 
$0\ell+3b$ & 11.9 (7.3)& 7.2 (4.3)  & 3.2 (1.9)& 0.9 (0.6)  & 1.2 (0.7) & 2.9 (1.7)\\ 
\hline 
$1\ell+3b$ & 8.6 (7.1)& 5.1 (4.2) & 2.8 (2.4) & 1.0 (0.8)& 0.9 (0.8) & 1.4 (1.2)\\
\hline 
$Z2b+V$ & 7.7 (7.4)& 2.2 (2.2)  & 1.2 (1.2)& 0.4 (0.4)& 0.6 (0.6)& 1.1 (1.1) \\ 
\hline  
$Z2b$ & 6.0 (5.7) & 2.3 (2.2) &0.8 (0.8) & 0.4 (0.4)& 0.5 (0.5) & 0.5 (0.5) \\
\hline  
$SS1b$ & 8.7 (6.6) & 5.3 (4.0) &2.9 (2.2)& 0.7 (0.6) & 0.6 (0.5)& 0.8 (0.6)\\ 
\hline  
$SS2b$ & 10.8 (9.9)& 7.1 (6.6) & 4.1 (3.8) & 1.2 (1.2)& 0.8 (0.8)& 1.3 (1.3)\\ 
\hline  
$\ell Zb$ & 3.5 (1.8)& 1.3 (0.7) & 0.6 (0.3) & 0.2 (0.1)& 0.2 (0.1)& 0.3 (0.2) \\ 
\hline  
$3\ell b$ & 6.3 (3.4)& 3.1 (1.7) & 1.6 (0.9) & 0.4 (0.2)& 0.3 (0.2)& 0.4 (0.2)\\ 
\hline 
Total & 23.5 (18.7)& 13.3 (10.4) &6.9 (5.6) & 2.1 (1.8) & 2.0 (1.6)& 3.7 (2.8)\\ 
\hline 
\end{tabular} 
\end{center}
\caption{The significances for a 300 fb$^{-1}$ integrated luminosity of Type A Benchmark points from various channels. The numbers in the brackets are significances with a $10\%$ background systematic uncertainty.}
\label{table:A_significance}
\end{table}

\begin{table}[ht]
\captionsetup{justification=raggedright,
singlelinecheck=false}
\begin{center}
\begin{tabular}{|c|c|c|c|c|}
\hline 
Significance & B1 & B2 & B3 & B4  \\ 
\hline 
$0\ell+3b$  & 5.9 (3.5) & 9.4 (5.6) & 3.5 (2.1) & 5.6 (3.3)\\ 
\hline 
$1\ell+3b$   & 4.1 (3.4) & 2.3 (2.0) & 1.6 (1.4) & 3.6 (3.0)\\
\hline 
$Z2b+V$  & 3.3 (3.2)& 1.8 (1.7)& 0.9 (0.9)& 1.0 (1.0)\\ 
\hline  
$Z2b$  & 3.1 (3.0) & 2.1 (2.0)& 0.6 (0.6)& 0.4 (0.4) \\
\hline  
$SS1b$   & 2.9 (2.2) & 0.2 (0.2)& 0.5 (0.4)& 0.7 (0.5)\\ 
\hline 
$SS2b$  & 4.7 (4.4) & 0.6 (0.6)& 1.3 (1.2)& 2.3 (2.1)\\ 
\hline  
$\ell Zb$  & 1.1 (0.6)& 0.5 (0.3)& 0.3 (0.2) & 0.2 (0.1)\\ 
\hline  
$3\ell b$  & 1.9 (1.0)& 0.2 (0.1)& 0.4 (0.2)& 0.5 (0.3)\\ 
\hline 
Total & 10.4 (8.3)& 10.1 (6.6)& 4.3 (3.0)& 7.2 (5.2) \\
\hline 
\end{tabular} 
\end{center}
\caption{The significances for a 300 fb$^{-1}$ integrated luminosity of Type B Benchmark points from various channels. The numbers in the brackets are  significances with a $10\%$ background systematic uncertainty.}
\label{table:B_significance}
\end{table}

Due to the unknown systematic uncertainties in different channels and possible improvements from dividing events into different signal regions, the numbers in Tables~\ref{table:A_significance} and \ref{table:B_significance} should not be taken literally. However, they provide a good guidance on the effectiveness of various search channels for different benchmark models. In Type A models, the same-sign dilepton signals are often the most effective channels due to low backgrounds and large fractions of final states containing multiple $W$'s. However, for A6, the channels with multi-$b$-jets have the best reach because the large decay branching fraction for $\ttwo \to h \tone$ and a small cross section for the heavy $\bone$. For Type B models, the longer decay chains through charginos and neutralinos often produce more $Z$, $h$, $t$ and $b$ in the final states, which results in more $b$'s, as can be seen from Table~\ref{table:B_final}. Therefore, the $0l3b$ may have the best reach. Most channels receive contributions from both $\ttwo$ and $\bone$. On the other hand, the $Z2b(V)$ and $\ell Zb$ channels are almost stop specific and receive little contribution from the sbottom. Due to the complicated decay patterns of $\ttwo$ and $\bone$, a combination of different search channels not only enhances the search reach, but also provides important information about the spectrum and the decay patterns upon the discovery by comparing signals in different channels.

\section{Conclusions}
\label{sec:conclusions}

If the hierarchy problem is solved by SUSY, the stops are likely to be light enough to be accessible at the LHC. On the other hand, to obtain a Higgs boson mass of 125 GeV, a large $A_t$ term in the stop sector is needed in MSSM to increase the radiative contribution if both stops are light. The large mixing between the left-handed and right-handed stops implies a large splitting ($\gtrsim 300$ GeV most of the time for stops lighter than $\sim 1$ TeV) between the two mass eigenstate and at least a sbottom lighter than the second stop. Such a stop and sbottom spectrum often embroil complex decay chains for the stop and sbottom sector.

Even though intensive LHC searches for the lightest stop have put strong constraints on its mass, it can be hidden in the compressed region where it is difficult to have an effective search. In that case, the second stop and the sbottom may be easier to discover. The existing experimental searches are based on the simplified model approach. In particular for the second stop, only $\ttwo \to \tone Z$ and and $\ttwo \to \tone h$ decays are assumed. We consider many benchmark models for a more complete stop and sbottom spectrum and find that the simplified models are seldom good approximations to the realistic models. The decay patterns are complex and often without a dominant mode. The branching ratios of $\ttwo \to \bone W$ and $\bone \to \tone W$ decays can be quite substantial and they were not considered in the existing experimental searches. If there are additional charginos and neutralinos lighter than $\ttwo$ or $\bone$, they can appear in the decay chains and make the decay pattern even more complex.

In this paper, we perform a study of collider searches of $\ttwo$ and $\bone$ at 14 TeV LHC assuming that $\tone$ is hidden in the compressed region. The study is based on general MSSM but focuses on the stop and sbottom sector. The spectra are divided into two types, depending whether there are additional charginos and neutralinos besides the LSP below the $\ttwo$ mass. We derive the branching ratios of various decay modes and obtain the fractions of possible final states. From there we can identify potentially useful signal channels for the $\ttwo$ and $\bone$ searches. In additional to the standard signals based on multi-$b$-jets and leptonic decaying $Z$'s of current experimental searches, we find that same-sign dilepton and multi-lepton signals are also important for $\ttwo$ and $\bone$ searches, because many benchmark models produce large fractions of multi-$W$ final states. The same-sign dilepton excesses observed in the LHC Run 1 and Run 2 data may be explained by some of our benchmark models if they turn out to be real. For the standard leptonic decaying $Z$, we find that additional vector-tags and a new kinematic variable which we called ``leverage'' can further help to increase signal significance.

Due to the complex decay patterns, which signal channels are most useful for $\ttwo$ and $\bone$ searches depend on the models and spectra. More often than not there is no dominant decay mode and a combination of many different signal channels is needed to obtain the best reach. Also, most signal channels receive contributions from both $\ttwo$ and $\bone$. Some signals with a reconstructed $Z$ may be thought as more $\ttwo$ specific. However, in Type B models, $Z$ can also appear in the $\bone$ decay chain from $\tilde{\chi}_2^0$ or  $\tilde{\chi}_3^0$ decay if they are lighter than $\bone$. It is therefore difficult to perform independent searches for $\ttwo$ and $\bone$ in a realistic scenario. Observation of signals in multiple channels and their kinematic distributions will be needed to help disentangling the ultimate underlying theory and its spectrum.

\section*{Acknowledgments}
We would like to thank Tim Cohen, Dmitri Denisov, and John Stupak for help with the Snowmass background simulations, also Thomas Hahn for FeynHiggs coding and Margarete M\"uhlleitner for the SUSYhit support. We also thank Ian Low and Angelo Monteux for discussion and email correspondences. H.-C. C. thanks Academia Sinica in Taiwan and Kavli Institute for Theoretical Physics China for hospitality where this manuscript is finished. This work is supported in part by the US Department of Energy grant DE-SC-000999.

\begin{appendix}

\section{Compatibilities of the benchmark models with current constraints from 13 TeV LHC data}
In this Appendix, we check our benchmark models against the experimental constraints from the most recent LHC 13 TeV Run 2 results.  For our analyses, the most relevant search channels from current LHC public results are multiple-$b$ jets with 0 or 1 lepton and same-sign dilepton with one or more $b$-jets. Other channels are either less significant such as multi-lepton channels, or not available yet such as the $Z2b$ channel. 

\subsection{Multiple $b$ + 0/1$\ell$}
For multiple $b$ + 0/1$\ell$ channels, the strongest constraints come from the searches of gluinos decaying via stop or sbottom~\cite{ATLAS:2016uzr}. In order to test the viability of our benchmark points, we adopt a cut similar to the ATLAS study:
All candidate jets should have $p_T>30$~GeV with $|\eta|<2.5$. All candidate leptons should have $p_T>20$~GeV with $|\eta|<2.5$. The 4 leading hardest jets are required to have a $\Delta\phi>0.4$ from the MET. At least 3 candidate jets should be $b$-tagged.
We define the inclusive effective invariant mass $M_{\text{eff}}$ to be the scalar sum of $p_T$ of all candidate jets and leptons and MET. The transverse mass of $b$ jets $M^b_{T,min}$ is defined to be $\text{Min}(\sqrt{2 p_{Tb} (1-\cos\Delta\phi_{i})}), i=1,2,3$, the minimum is taking among three leading $b$ jets. Finally, we define the total jet mass variable $M^{\Sigma}_J$ to be the mass sum of the 4 leading fat jets with $p_T>100$~GeV. Five signal regions used to search for $\tilde{g}\to \tilde{t}t$ are given by:
\begin{table}[h]
\begin{tabular}{|c|c|c|}
\hline
Name & Definition & ATLAS 95$\%$ CL limit\\
\hline
Gtt0LA & $N_j \geqslant$ 8, $M^b_{T,min}>80$, MET$>$400, $M_J^\Sigma>200$, $M_{\text{eff}}>2000$       &3.8\\
\hline
Gtt0LB & $N_j \geqslant$ 8, $M^b_{T,min}>80$, MET$>$400, $M_{\text{eff}}>1250$       &13.3\\
\hline
Gtt1LA & $N_j \geqslant$ 6, $M^b_{T,min}>120$, $M_T>200$, MET$>$200, $M_J^\Sigma>200$, $M_{\text{eff}}>2000$   &3.8\\
\hline
Gtt1LB & $N_j \geqslant$ 6, $M^b_{T,min}>120$, $M_T>200$, MET$>$350, $M_J^\Sigma>150$, $M_{\text{eff}}>1500$   &4.9\\
\hline
Gtt1LA & $N_j \geqslant$ 6, $N_{jb} \geqslant$ 4, $M^b_{T,min}>80$, $M_T>150$, MET$>$200, $M_{\text{eff}}>500$   &5.7\\
\hline
\end{tabular}
\end{table}

These are then applied upon our benchmark points. For A1, we expect 0.4 and 0.9 events for two 0L signal regions. For 1L + multi-$b$ channels Gtt1L-A/B and C, A1 gives 0.1, 0.3 and 0.5 signal events with 14.8 $\text{fb}^{-1}$. All of them are an order of magnitude less than the 95\% CL exclusion limit. The next lightest benchmark point A2 contributes 0.2, 0.7, 0.03, 0.08 and 0.2 to these 5 signal regions. Other Type A benchmark points yield smaller numbers. For Type B models, B1 would contribute 0.2 and 0.5 events to two 0L signal regions, and 0.06, 0.2, 0.3 events for the 1L regions. The contributions from other type B benchmark points are even less. The smallness of the number of events is partly due to the hard cuts imposed in the analyses which are designed for the gluino search. Therefore, none of our benchmark points are excluded by the multi-$b$ +0/1 lepton searches.

\subsection{Same-sign dilepton +$b$ jets}
For this signal we compare to the ATLAS same-sign dilepton SUSY search~\cite{ATLAS:2016kjm}. The most relevant signal region is SR1b and SR3b. We require each candidate jet must have $p_T>25$~GeV and $|\eta|<2.5$. Each candidate lepton need to have $p_T>10$~GeV and $|\eta|<2.5$. The signal requires a pair of same-sign lepton or more than 3 leptons. 
\begin{table}[h]
\begin{tabular}{|c|c|c|}
\hline
Name & Definition & ATLAS 95$\%$ CL limit\\
\hline
SR1b & $N_j \geqslant$ 6, $N_b\geqslant1$, MET$>$200,  $M_{\text{eff}}>650$       &10.3\\
\hline
SR3b & $N_j \geqslant$ 6, $N_b\geqslant3$, MET$>$160, $M_{\text{eff}}>600$       &4.9\\
\hline
\end{tabular}
\end{table}
We find that at 13~TeV and the same luminosity, the lightest benchmark A1 gives 8.0 events for SR1b and 0.9 events for SR3b, thus is not excluded. Other benchmark points produce even less events. For instance, A2 gives 4.9 and 0.4 for these two signal regions while B1 gives 2.5 and 0.2. Note that with more data, A1 might be excluded soon, while testing other benchmark points may take longer time because their contributions are well below the upper limits.

\subsection{CMS jets+MET data}

The models points can also be constrained by SUSY searches with multi-jets and MET. In particular, CMS showed a deficit in the aggregate search region SR11 in ``Additional Table 5'' of Ref.~\cite{CMS:2016mwj}. This aggregate region is targeted for a light SUSY spectrum. It requires to 7+ jets with 1+ b jets, each of them should have $p_T>30$~GeV, $H_T \geq 300$~GeV and $H_T^{\rm miss}\geq 300$~GeV. Also no isolated leptons or isolated-charged particles are found. For the two leading jets, their $\Delta\phi$ with the MET are required to be greater than 0.5, while for the third and the fourth leading jets, $\Delta\phi>0.3$ is required. The predicted number of background events is $385^{+19+27}_{-17-27}$, while the observed number is 316, showing a $\sim 2\sigma$ deficit. This can put any new physics models that contribute a significant number of events in this region in tension with the data.

As a test of our benchmark models, the lightest A1 point gives 35 events for SR11, which roughly equals the $1\sigma$ uncertainty of the predicted background events, and A2 contributes about 22 events. They are in some tension with the observed data, but not much more than the SM itself. We note that among the 12 combined signal regions, SR11 is the only one showing a significant deficit. We hope that with more statistics and a cross-check with ATLAS will clarify this situation soon.

\end{appendix}



\end{document}